\documentclass{aa}

\usepackage{graphicx}
\usepackage{txfonts}
\usepackage{soul}
\usepackage{booktabs}
\usepackage{multirow}
\usepackage{xcolor}
\begin{document}

\title{Confirmation of the planetary nebula nature of HaTr\,5}
\subtitle{Not the remnant of Nova Sco\,1437}

\author{
M.~A.~Guerrero
\inst{1}
  \and
E.~Santamar\'\i a
\inst{2}
  \and
G.~Liberato
\inst{3}
  \and
Q.~A.~Parker
\inst{4}
  \and
D.~R.~Gon\c{c}alves
\inst{3}
\and
J.~B.~Rodr\'\i guez-Gonz\'alez
\inst{2}
  \and
A.~Ritter
\inst{4}
  \and
H.~Yuan
\inst{4}
  \and
J.~A.~Toal\'a
\inst{2}
}

\institute{
Instituto de Astrof\'{i}sica de Andaluc\'{i}a, IAA-CSIC, Glorieta de la Astronom\'{i}a S/N, Granada 18008, Spain\\
\email{mar@iaa.es}
  \and
Instituto de Radioastronom\'{i}a y Astrof\'{i}sica, Universidad Nacional Aut\'{o}noma de M\'{e}xico, 58090 Morelia, Mich., Mexico
  \and
Observat\'{o}rio do Valongo, Universidade Federal do Rio de Janeiro, Ladeira Pedro Antonio 43, Rio de Janeiro 20080-090, Brazil
  \and
Laboratory for Space Research, The University of Hong Kong
}

\date{\today}

\abstract{
The identification of the nebula HaTr\,5 with the shell remnant of the historic Nova Sco\,1437 around the low-accretion rate cataclysmic variable 2MASS\,J17022815-4306123 has been used in the framework of the hibernation scenario to set an upper limit of $\leq$580 yr to the transition time from a nova-like binary to a dwarf nova.  
}{
This work aims at clarifying the nature of HaTr\,5, which has also previously been proposed to be a possible planetary nebula.
}{
Intermediate- and high-dispersion long-slit spectra of HaTr\,5 have been obtained and analyzed in conjunction with 
archival optical and infrared images to investigate its spectral properties using photoionization models, to derive its 
H$\alpha$ flux and ionized mass, and to determine its spatio-kinematic by means of 3D models to clarify its true nature. 
}{
The H$\alpha$ flux of HaTr\,5 implies an ionized mass of 0.059 M$_\odot$ at the 0.99 kpc distance of 2MASS\,J17022815-4306123, i.e., $\sim$1000 times the typical ejecta of a nova.  
If HaTr\,5 were actually an unrelated planetary nebula, its H$\alpha$ flux implies a distance of 2.25 kpc and an ionized mass of 0.47 M$_\odot$. 
The expansion velocity of HaTr\,5 is found to be $\simeq$27 km~s$^{-1}$, with a heliocentric radial velocity of 
$-1$ km~s$^{-1}$. 
}{
The ionized mass of HaTr\,5 and its expansion velocity (and associated kinematic age) are clearly inconsistent with those expected for a nova remnant, which all strongly support a planetary nebula nature. 
The association of 2MASS\,J17022815-4306123 with HaTr\,5 is further called into question by their differing radial velocities and almost orthogonal motions on the plane of the sky. 
It is concluded that HaTr\,5 is an old, evolved planetary nebula unrelated to the remnant of Nova Sco~1437 and to the cataclysmic variable 2MASS\,J17022815-4306123, the latter being by chance projected onto HaTr\,5.  
}


\maketitle
    

\section{Introduction}
\label{sec:intro}

Ionized gaseous nebulae in our Galaxy arise from multiple processes associated with different progenitors that can have 
vastly different origins, masses and physical states on observation. 
They range from H~{\sc ii} regions, ionized bubbles, including Str\"omgren spheres around hot white dwarf (WD) stars or 
massive stars, and various kinds of stellar ejecta at different phases of evolution such as Wolf-Rayet shells, nova
remnants, supernova remnants, symbiotic systems, Herbig-Haro outflows, and planetary nebulae (PNe hereafter). 
A variety of powerful diagnostic tools are used to discriminate among them, including multi-wavelength imagery, kinematic 
properties, spectroscopy, emission line ratios, and even the local environment. 
Many nebula types have properties in common, however, and sometimes it is not straightforward to be equivocal about their nature, even when a holistic approach to consideration of all available source and comparison data are used 
\citep[see, e.g.,][where various PN mimics are discussed]{FP2010,Parker2022}.

Here we review a specific and interesting contentious source, the nebula HaTr\,5, whose true nature is disputed between PN and nova remnant.

\subsection{The nature of the nebula HaTr\,5}
\label{sec:HaTr5}

HaTr\,5 is a flocculent, approximately round nebula proposed to be a low surface brightness PN by its discoverers \citep{HaTr1985}.  
Its optical spectrum, while ruling it out as a normal H~{\sc ii} region due to the high observed [N~{\sc ii}] to H$\alpha$ line ratio, is not typical of a PN \citep[e.g.,][]{FP2010,Parker2022}, because of its very high [S~{\sc ii}] $\lambda6716,6731$ to H$\alpha$ line ratio, $\sim1.0$, which is most usually seen in sources with strong shock-excitation, such as supernova remnants or Wolf-Rayet shells. 
This fact, and the absence of any obvious central star, have led to its classification as originally only a possible PN in the most comprehensive and robust available catalog of PNe, the Hong Kong/AAO/Strasbourg H-alpha planetary nebula database \citep[HASH,][]{Parker+2016}\footnote{\url{http://202.189.117.101:8999/gpne/}}, where it has HASH ID number 1105.

\begin{table*}
\caption{Spectroscopic observation log of HaTr\,5} 
\label{tbl:ObsLog}
\centering
\begin{tabular}{llllrrr}
\hline\hline
\multicolumn{1}{l}{Telescope} & 
\multicolumn{1}{l}{Spectrograph} &  
\multicolumn{1}{l}{Date} & 
\multicolumn{1}{l}{Object} & 
\multicolumn{1}{c}{$t_\mathrm{exp}$} & 
\multicolumn{1}{c}{PA} & 
\multicolumn{1}{c}{Slit Width} \\
\multicolumn{4}{c}{} & 
\multicolumn{1}{c}{(s)} & 
\multicolumn{1}{c}{($^{\circ}$)} & 
\multicolumn{1}{c}{($^{\prime\prime}$)} \\
\hline
SOAR 4.1m & GHTS     & July 1, 2024  & HaTr\,5 & 3$\times$1,800 &  45~~~ & 0.6~~~~~~ \\
SOAR 4.1m & GHTS     & July 1,  2024 & HaTr\,5 & 3$\times$1,800 & 135~~~ & 0.6~~~~~~ \\
SAAO 1.9m & CassSpec & September 15, 2014 & HaTr\,5 &       600 &  90~~~ & 2.0~~~~~~ \\
SAAO 1.9m & SpUpNIC  & June 8, 2024 & HaTr\,5 &  3$\times$1,200 &  90~~~ & 2.0~~~~~~ \\
SAAO 1.9m & SpUpNIC  & June 8, 2024 & HaTr\,5 sky &       1,200 &  90~~~ & 2.0~~~~~~ \\
\hline
\end{tabular}
\end{table*}

HaTr~5 has also been proposed to be associated with the gaseous remnant of the historical nova Scorpii of 11 March {\sc ad} 1437, hereafter Nova Sco\,1437 \citep{Shara+2017}. 
The association is based on the identification of the putative nova progenitor with the cataclysmic variable (CV) 2MASS\,J17012815-4306123 and X-ray source IGR\,J17014-4306 (hereafter J17014-4306), which is projected onto HaTr\,5 and whose proper motion placed it at its center at the time of the nova event.

J17014-4306 is a dwarf nova with small accretion rate. 
Therefore its identification as the progenitor of Nova Sco\,1437 is relevant for the hibernation hypothesis, which predicts that the mass-transfer rate onto the WD exhibits a cyclic behavior, remaining high for centuries after the nova eruption to then become low for thousand to million years before a new nova eruption occurs \citep{Shara+1986}. 
Accordingly a classical nova (CN) is expected to become a nova-like binary after the eruption to turn a few centuries later into a dwarf nova.  
There are a few examples of ancient nova shells around dwarf novae \citep[e.g., Z~Cam,][]{Shara+2007} that would support these expectations, but most lack a robust determination of the nova date.  
This makes the case of Nova Sco\,1437 potentially unique, as it provides one of the few examples of a dwarf nova associated with a CN remnant of known age.

The hibernation hypothesis is appealing and has important implications for stellar astrophysics, as it settles the mass grow rate of WDs in accreting binary systems in their possible path towards Type~Ia supernova progenitors \citep[as proposed for SN Ia PTF 11kx,][]{Dilday+2012}.   
Observational evidence of the time-lag from the high to low accretion rate states is, however, controversial.  
A key system is WY\,Sge, recorded as the oldest CN eruption in the late 18th century (Nova WY Sag\,1783). 
It exhibits nowadays a high mass loss rate transfer to the WD \citep{Somers+1996,Vogt+2018}, thus providing a lower limit of $\geq$240~yr for the transition from nova-like systems to dwarf novae \citep{Shara1984}. 
This is also the case of V1213\,Cen (a.k.a.\ Nova Cen\,2019), whose continuous photometric coverage revealed a dwarf-nova (low accretion rate) before the nova eruption, but a nova-like binary (high accretion rate) afterwards \citep{Mroz+2016}. 
On the contrary, there are quite a number of examples of post-novae with dwarf-nova like outbursts just a few decades \citep[e.g.\ V446\,Her, a.k.a.\ Nova Her\,1960,][]{HRK2011} or in the range from one hundred to two hundreds years \citep[e.g., V728\,Sco, a.k.a.\ Nova Sco\,1862,][]{Tappert+2012} after the nova eruption. 
Most likely the time-lag before the hibernation phase starts depends on the specifics of the binary system \citep{Simon2023}.

Being Nova Sco\,1437 relevant to the discussion on the transition from high to low accretion rate states after a nova eruption, the identification of its remnant with HaTr\,5  has been questioned after a critical revision of 
ancient Chinese and Korean astronomy texts \citep{Hoffmann2019}. 
This was motivated by a reduction to about a third of the previous proper-motion of J17014-4306 by Gaia DR2 data \citep[confirmed in GAIA DR3 data,][]{Bailer-Jones+2021}. 
The age of the remnant of Nova Sco\,1437 is also unusually large compared to most nova remnants, whose H$\alpha$ luminosity declines below typical detection limits before they are 200 yr old \citep{Tappert+2020} with the exception of a few ancient nova remnants proposed to be several centuries old, e.g., V1315\,Aql, BZ\,Cam, Te\,11 \citep{Miszalski+2016,Sahman+2018,BM2018}, and others in addition to Z\,Cam \citep[see the review on ancient novae in][]{Tappert+2020}.

In this paper, we present new, long-slit intermediate- and high-dispersion spectroscopic observations of HaTr~5 to 
investigate afresh the properties of this nebula. Focus is placed on the comparison of the nebular kinematics 
expected for a nova remnant and a PN as a key diagnostic.  
An expansion velocity of a few hundreds km~s$^{-1}$ would point to a nova remnant.  
An expansion of a few tens km~s$^{-1}$ would be attributable to a PN. 
Finally, a negligible expansion velocity would be associated with a Str\"omgren sphere around a WD, as long-lived WDs, whose own PN dissipated in the distant past, migrate into an area of ISM that gets ionized by the UV radiation field of such hot stars.

This paper is organized as follows. 
Observations are described in Section \ref{sec:obs}. 
The nebula properties and 
consequences of our spectroscopic observations are presented in Section~\ref{sec:prop} while our discussion of the alternative identification scenarios are given in Section~\ref{sec:disc}.
Final remarks on the possible association between HaTr~5 and the cataclysmic variable J17014-4306 are given in 
Section~\ref{sec:remarks}. 

\section{Observations and data reduction} 
\label{sec:obs}

\subsection{Archival VPHAS+, SHS and Spitzer images}

\begin{figure*}
\centering
\includegraphics[angle=0,height=0.45\linewidth]{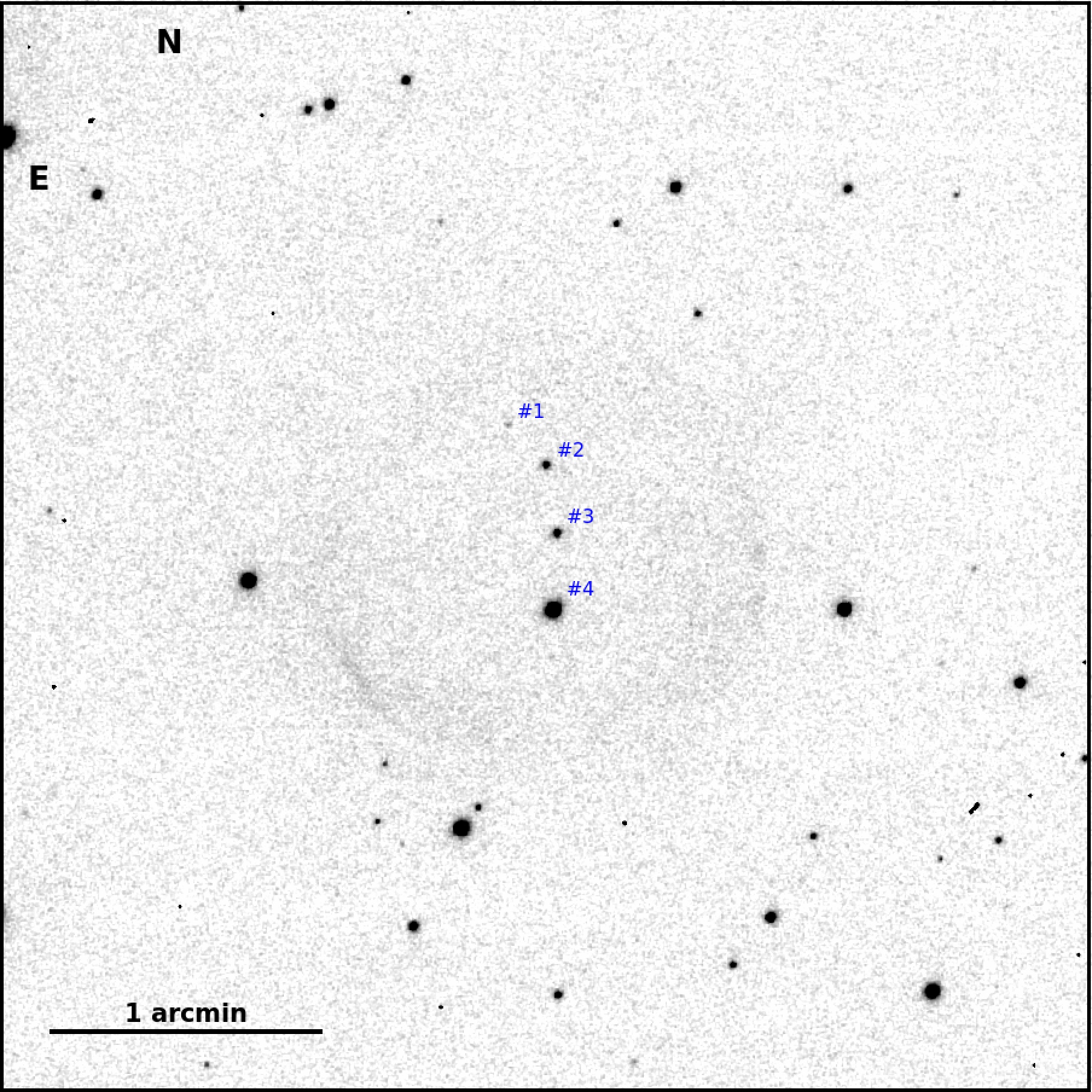}
\,
\includegraphics[angle=0,height=0.45\linewidth]{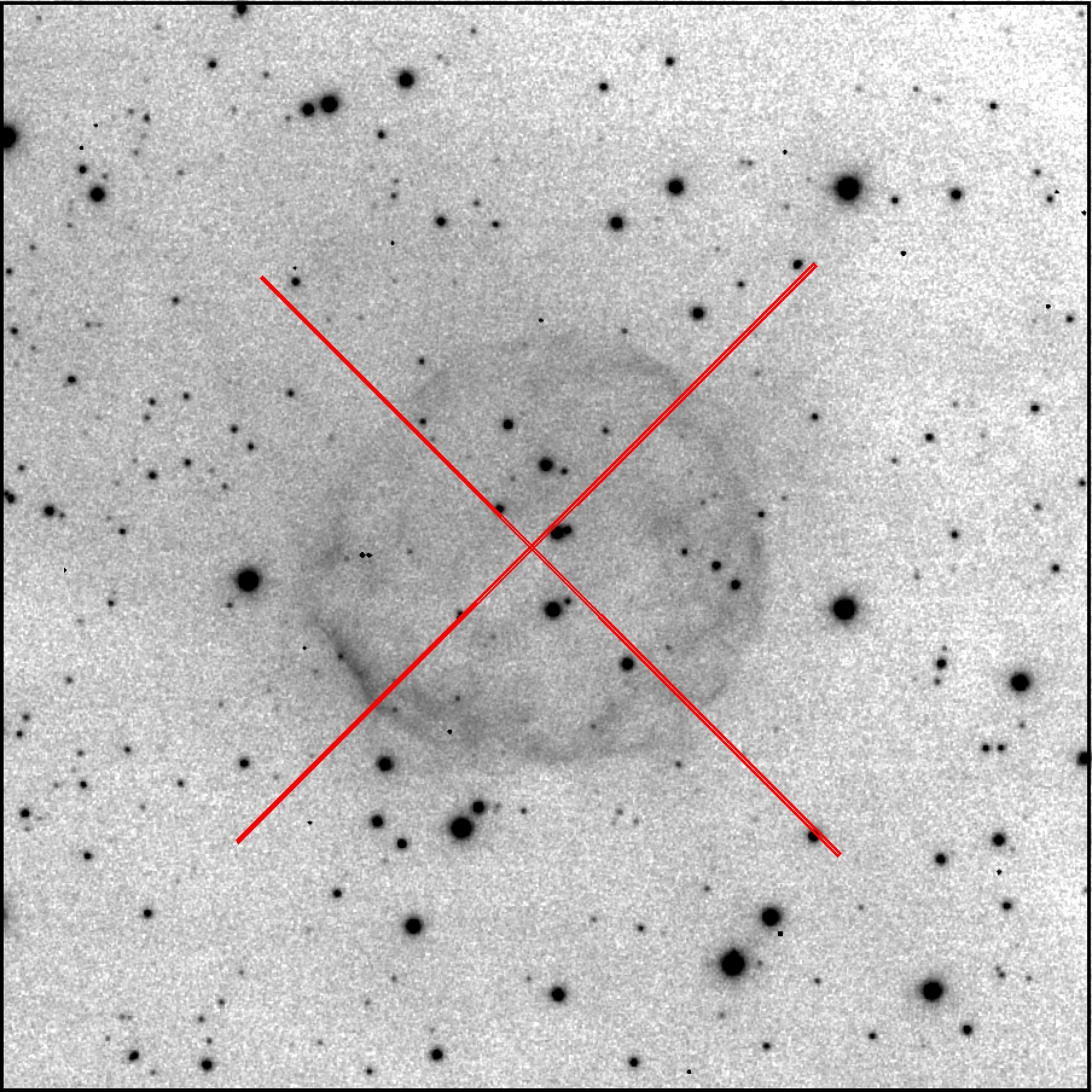}
\\
\vspace*{0.2cm}
\includegraphics[angle=0,height=0.45\linewidth]{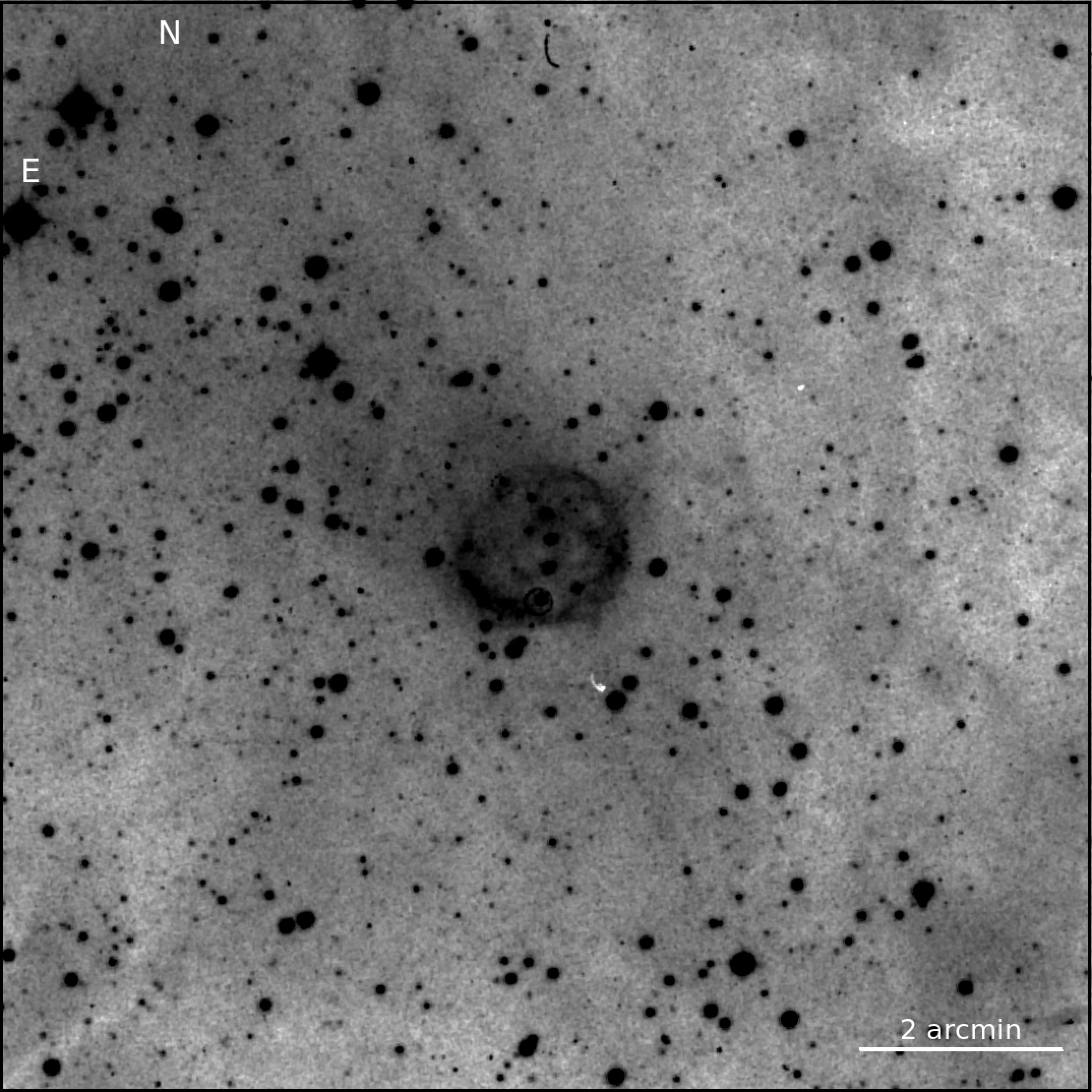}
\,
\includegraphics[angle=0,height=0.45\linewidth]{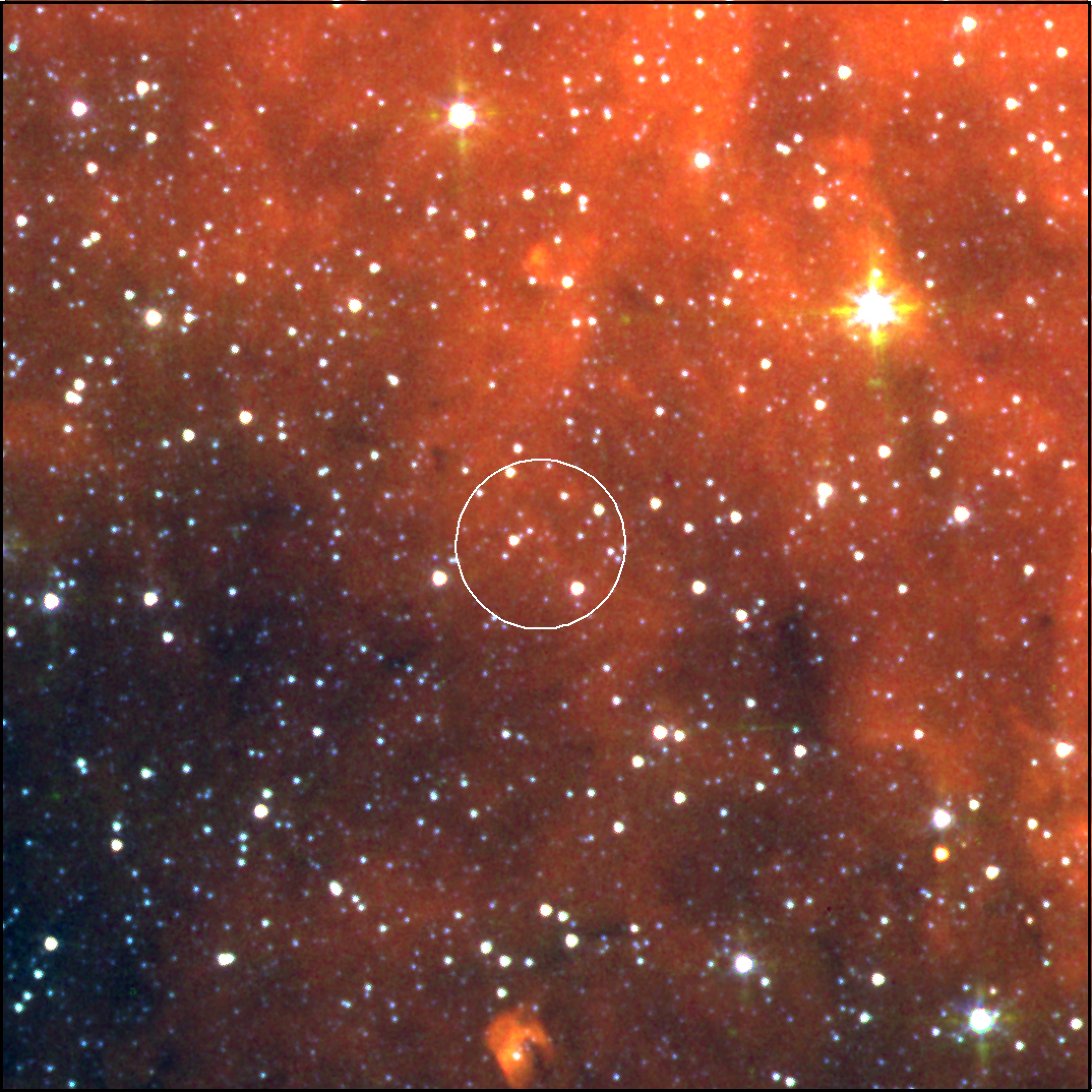}
\caption{\small 
VPHAS+ $u$-SDSS (top-left) and H$\alpha$+[N~{\sc ii}] (top-right) images centered on HaTr\,5 (field of view $\approx$4 
arcmin), and wide field of view ($\approx$10.5 arcmin) SHS H$\alpha$+[N~{\sc ii}] image (bottom-left) and Spitzer IRAC color-composite picture (bottom-right) of HaTr\,5.  
Stars detected in the $u$ band projected onto HaTr\,5 are labeled on the $u$-SDSS image, whereas the location and orientation of the two SOAR GHTS long-slits are overlaid on the H$\alpha$ image.  
On the Spitzer IRAC picture, red, green, and blue colors correspond to the IRAC4 8 $\mu$m, IRAC3 5.8 $\mu$m, and IRAC2 
4.5 $\mu$m bands, respectively, whereas the location of the nebula is marked by a white circle. 
The wide field of view images in the bottom panels highlight the general diffuse emission around HaTr\,5, with no obvious 
connection between the mid-IR and optical nebular emissions. }
\label{fig:img}
\end{figure*}

Images in the $u$, $g$, and H$\alpha$ band of the sky region including HaTr\,5 were downloaded from the VST Photometric 
H$\alpha$ Survey of the Southern Galactic Plane and Bulge \citep[VPHAS+,][]{Drew+2014} archive.  
VPHAS+ uses the 2.6~m VLT Survey Telescope (VST) on ESO's Cerro Paranal Observatory and the large field-of-view OmegaCAM.  
The images have a pixel size of 0.213 arcsec.  
The $u$ and $g$ band images were obtained on September 11 2014 using $u$-SDSS and $g$-
SDSS filters with total exposure times of 150~s and 40~s, respectively. 
The H$\alpha$ image was obtained on July 24 2012 using the NB$\_$659 filter ($\lambda_\mathrm{c}=6588$ \AA, FWHM = 107 
\AA) with a total exposure time of 120~s. The filter bandpass includes emission both from the H$\alpha$ and [N~{\sc 
ii}] $\lambda\lambda$6548,6584 emission lines. The resulting image will be accordingly referred to as H$\alpha$+[N~{\sc ii}] 
image.  The $u$ band and H$\alpha$+[N~{\sc ii}] images are presented in the top-left and top-right panels of 
Fig.~\ref{fig:img}, respectively.

A wide field of view SuperCOSMOS H$\alpha$ Survey \citep[SHS,][]{Parker+2005} H$\alpha$+[N~{\sc ii}] image including 
HaTr\,5 was downloaded from the SHS Homepage\footnote{\url{http://www-wfau.roe.ac.uk/sss/halpha/}}.  
The image, shown in the bottom-left panel of Fig.~\ref{fig:img}, has a pixel size of 0.677 arcsec.  

Similar wide field of view Spitzer IRAC images in the IRAC2 4.5$\mu$m, IRAC3 5.8$\mu$m, and IRAC4 8.0 $\mu$m bands were 
downloaded from the IRSA Archive\footnote{\url{https://irsa.ipac.caltech.edu/frontpage/}}.  
These images have been used to made the IRAC color-composite picture presented in the bottom-right panel of 
Fig.~\ref{fig:img}.  This picture has the same field-of-view of the SHS image presented in the same figure.

\subsection{High-dispersion spectroscopy}

The Goodman High Throughput Spectrograph (GHTS) at the 4.1~m Southern Astrophysical Research Telescope (SOAR) on the Cerro 
Pach\'on (Chile) was used to obtain high-resolution long-slit spectra of HaTr\,5 on July 1 2024.  
The red camera and 2100 lines~mm$^{-1}$ grating with central wavelength 6507~\AA\ were used to cover the spectral 
range from 6130~\AA\ to 6710~\AA. 
This includes the bright H$\alpha$ and [N~{\sc ii}] $\lambda\lambda$6548,6584~\AA\ and the low-excitation [O~{\sc i}] $\lambda\lambda$6300,6363\AA\ emission lines prominent in the night sky spectrum and in supernova remnants but typically absent or weak in PN spectra.  
The CCD e2V CCD231-84-1-F21 was used, providing a spectral dispersion of 0.141 \AA~pix$^{-1}$ and a plate scale of 0.15 arcsec~pix$^{-1}$.  
In conjunction with the 0.6$^{\prime\prime}$ wide slit, this setup provides a spectral resolution of $\simeq$10,000, i.e., $\simeq$30--40 km~s$^{-1}$ for the above emission lines. 

As stated in the observation log detailed in Table~\ref{tbl:ObsLog}, three 1,800~s exposures were 
obtained for HaTr\,5 with slits chosen along two different position angles (PA), 
$45^\circ$ and $135^\circ$, as illustrated in the top-right panel of Fig.~\ref{fig:img}. 
The two PAs allow to make an assessment of the nebula kinematics, effectively removing the diffuse background emission 
surrounding HaTr\,5 (see the IR image in the right panel of Fig~\ref{fig:img}), which can affect line ratios and 
kinematics information as well. 

Bias and flat-field frames were acquired, as well as Hg+Ar+Ne arc lamp frames for wavelength calibration and spectral 
observations of the spectro-photometric standard star EGGR\,274 of WD spectral type for flux calibration 
\citep{Greenstein1984}.  The arc exposures were obtained immediately before the six exposures of HaTr~5, 
while those of EGGR\,274 were taken after. The data were reduced using standard IRAF routines \citep{Tody1993}. 
As determined from the FWHM of the continuum emission of stars 
registered along the slit, the spatial resolution (seeing) is estimated to be 
$\simeq$1.4 arcsec. 
The wavelength calibration accuracy is better than 0.1 km~s$^{-1}$.

\subsection{Intermediate-dispersion spectroscopy}

Two sets of intermediate-dispersion long-slit spectroscopic observations of HaTr\,5 were obtained with the South-African 
Astronomical Observatory (SAAO) 1.9~m telescope 10 years apart. Both sets of spectra used the 300 lines~mm$^{-1}$ grating \#gr7 and a 2~arsec wide slit oriented East--West, that is, a PA=90$^\circ$.

\begin{figure*}
\centering
\includegraphics[bb=1 145 560 490,angle=0,width=0.90\linewidth]{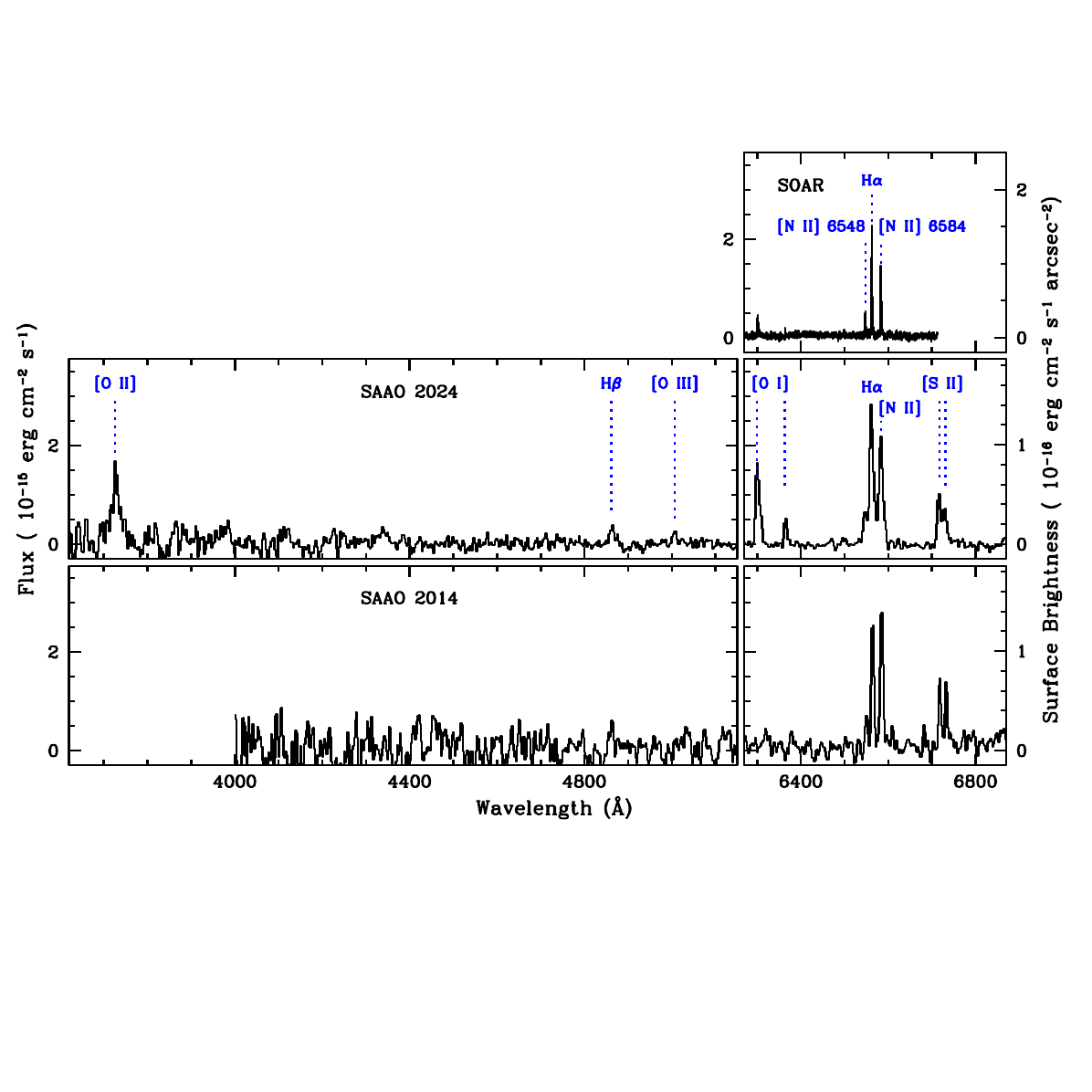}
\caption{\small 
SOAR (top), SAAO 2024 (middle), and SAAO 2014 (bottom) one-dimensional spectra of HaTr\,5. 
Emission lines are labeled on the SOAR and SAAO 2024 spectra. 
The left y-axis denotes flux, but surface brightness for the right y-axis.  
Note the ``artificial'' extreme variations of the [O~{\sc i}] emission lines caused by poor sky subtraction for the SAAO 2024 spectrum and low signal-to-noise ratio of the SAAO 2014 spectrum.   
The apparent [O~{\sc i}] emission lines in the SOAR spectrum are residuals of the sky subtraction as well. 
}
\label{fig:spec}
\end{figure*}

\begin{table*}
\caption{Measured line fluxes ($F$) and extinction corrected line intensities ($I$) of HaTr\,5, together with synthetic line intensities.
}
\label{tbl:spec_low}
\centering
\begin{tabular}{lrccccccc}
\hline\hline
 & 
 & 
\multicolumn{2}{c}{SAAO 2014} &  
\multicolumn{2}{c}{SAAO 2024} &
\multicolumn{2}{c}{\citet{Shara+2017}} &  
\multicolumn{1}{c}{Model} \\ 
\multicolumn{1}{l}{Line} & 
\multicolumn{1}{c}{$f(\lambda)$} & 
\multicolumn{1}{c}{$F/F(\mathrm{H}\alpha\mathrm)$} &
\multicolumn{1}{c}{$I/I(\mathrm{H}\alpha\mathrm)$} & 
\multicolumn{1}{c}{$F/F(\mathrm{H}\alpha\mathrm)$} &
\multicolumn{1}{c}{$I/I(\mathrm{H}\alpha\mathrm)$} & 
\multicolumn{1}{c}{$F/F(\mathrm{H}\alpha\mathrm)$} &
\multicolumn{1}{c}{$I/I(\mathrm{H}\alpha\mathrm)$} & 
\multicolumn{1}{c}{$I/I(\mathrm{H}\alpha\mathrm)$} \\
\hline
$[$O~{\sc ii}] 3726  & $+0.26$ & $\dots$         & $\dots$ & 0.55$\pm$0.19 : & 1.68 :   & $\dots$ & $\dots$ & 1.02 \\
H$\delta$ 4101       & $+0.18$ & $\dots$         & $\dots$ & $\dots$         & $\dots$ & 0.03    & 0.08    & 0.09 \\
H$\gamma$ 4340       & $+0.14$ & $\dots$         & $\dots$ & $\dots$         & $\dots$ & 0.05    & 0.12    & 0.17 \\
He~{\sc ii} 4686     & $+0.05$ & $\dots$         & $\dots$ & $\dots$         & $\dots$ & 0.017   & 0.035   & 0.02 \\
H$\beta$ 4861        &   0.00  & 0.21$\pm$0.07 : & 0.40 :  & 0.18$\pm$0.04   & 0.34    & 0.19    & 0.36    & 0.36 \\
$[$O~{\sc iii}] 5007 & $-0.03$ & $\dots$         & $\dots$ & 0.11$\pm$0.02   & 0.20    & 0.13    & 0.23    & 0.22 \\
$[$N~{\sc ii}] 5755  & $-0.21$ & $\dots$         & $\dots$ & $\dots$         & $\dots$ & 0.015   & 0.02    & 0.01 \\
He~{\sc i} 5876      & $-0.23$ & $\dots$         & $\dots$ & $\dots$         & $\dots$ & 0.03    & 0.02    & 0.01 \\
$[$O~{\sc i}] 6300   & $-0.30$ & $\dots$         & $\dots$ & 0.61$\pm$0.02 : & 0.66 :   & 0.17    & 0.19    & 0.10 \\
$[$O~{\sc i}] 6363   & $-0.30$ & $\dots$         & $\dots$ & 0.15$\pm$0.02 : & 0.16 :   & 0.06    & 0.07    & 0.03 \\
$[$N~{\sc ii}] 6548  & $-0.34$ & 0.26$\pm$0.06   & 0.26    & 0.25$\pm$0.02   & 0.25    & 0.35    & 0.36    & 0.31 \\
H$\alpha$ 6563       & $-0.34$ & 1.00$\pm$0.06   & 1.00    & 1.00$\pm$0.02   & 1.00    & 1.00    & 1.00    & 1.00 \\
$[$N~{\sc ii}] 6584  & $-0.34$ & 1.14$\pm$0.07   & 1.14    & 0.81$\pm$0.02   & 0.81    & 1.19    & 1.22    & 0.92 \\
$[$S~{\sc ii}] 6716  & $-0.36$ & 0.55$\pm$0.06   & 0.53    & 0.37$\pm$0.02   & 0.36    & 0.30    & 0.30    & 0.33 \\
$[$S~{\sc ii}] 6731  & $-0.36$ & 0.49$\pm$0.06   & 0.47    & 0.26$\pm$0.02   & 0.25    & 0.22    & 0.22    & 0.24 \\
\hline
$F({\rm H}\alpha)^a$  & & $2.0\times10^{-14}$ & & $3.1\times10^{-14}$ & & $\dots$ & &$\dots$ \\
SB(H$\alpha)^a$ & & $1.0\times10^{-16}$ & & $2.0\times10^{-16}$ & & $1.3\times10^{-16}$ & & $3.0\times10^{-16}$ \\
\hline
\end{tabular}
\tablefoot{
The line fluxes and intensities are relative to H$\alpha$ (=1.0).  
They correspond to observations obtained at SAAO on 2014 and 2024, and those from \citet{Shara+2017}, while the line intensity predictions have been derived from a Cloudy photoionization model.  
The observed H$\alpha$ flux $F({\rm H}\alpha)$ and surface brightness $SB$(H$\alpha$) are given in units of 
erg~cm$^{-2}$~s$^{-1}$ and erg~cm$^{-2}$~s$^{-1}$~arcsec$^{-2}$, respectively. 
Line fluxes and intensities affected by large uncertainties are denoted by a ":" sign.  
}
\end{table*}

The 2014 SAAO 1.9m data were obtained on September 15 with the old Cassegrain spectrograph and covered a wavelength range of 
4000~\AA~ to 7550~\AA~ with a spectral dispersion of 3.84~\AA~pix$^{-1}$. 
The usual bias, dome flat field, sky flat field, and Cu+Ar arc calibration frames were acquired. For the data reduction standard 
IRAF techniques were employed to deliver 1-D wavelength calibrated, sky subtracted spectra.

The 2024 data  were obtained on June 8 with the so-called "Spectrograph Upgrade–Newly Improved Cassegrain spectrograph" 
\citep[SpUpNIC,][]{Crause+2019} that covered a wider wavelength range of 5500~\AA, from 3500~\AA\ further out to the blue and to the 
red at 9000~\AA\ due to the different format CCD and new spectrograph design.
The spectral dispersion provided was 2.7 \AA~pix$^{-1}$. Once more, standard bias, dome flat field, sky flat field, and Cu+Ar 
arc frames were acquired.  Arc spectra were taken immediately before and after the science observations in both data sets. 

For the 2024 data reduction we used standard IRAF techniques coupled with our in-house Python-based data-reduction pipeline. 
This pipeline performed the standard reduction steps like overscan subtraction, trimming, bias subtraction, cosmic-ray removal 
with the L.A.Cosmic algorithm by \cite{vanDokkum2001}, removing pixel-to-pixel sensitivity variations using the master dome flat 
(ignoring pixels with a S/N less than 100), and correcting for the CCD illumination profile using the master sky flat. 
Before the extraction of the spectra to 1-D using the simple-sum algorithm, the 2-D spectra were transformed so that the 
spectral features align with columns and rows. In each case the sky was subtracted using a separate sky spectrum close to 
HaTr\,5. Unfortunately, due to bad weather, the sky spectrum had to be taken the following night (September 9, 2024) at a 
similar time. This likely accounts for the presence of residual [O~{\sc i}] 6300,6363 night sky-lines in the 2024 data.
After the wavelength calibration (RMS $<$0.4 km s$^{-1}$) the spectra were corrected for the heliocentric velocity using the 
Astropy SkyCoord package and flux calibrated using the spectrophotometric standard star HR 7950 with the Astropy 
Specreduce package.

\begin{figure*}
\centering
\includegraphics[angle=0,width=1.0\linewidth]{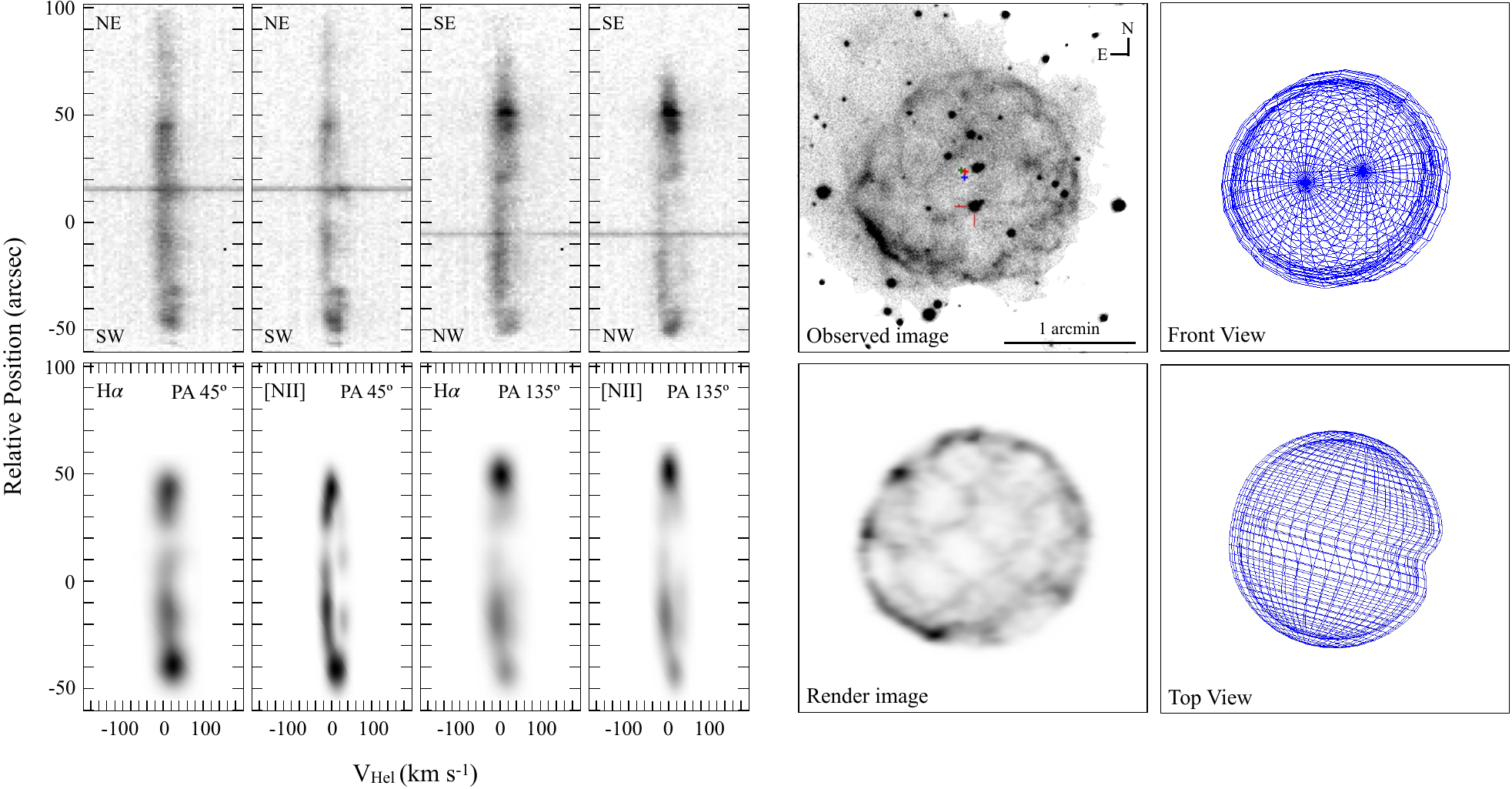}
\caption{\small 
Spatio-kinematic information of HaTr\,5 and the best SHAPE model.  
(left) SOAR GHTS H$\alpha$ and [N~{\sc ii}] (top) and SHAPE synthetic (bottom) PV diagrams. 
(center) HASH H$\alpha$+[N~{\sc ii}] (top) and {\sc shape} synthetic (bottom) image.  
(right) SHAPE mesh models along the line of sight and from the plane of the sky along the Southeast to Northwest direction.  
}
\label{fig:shape}
\end{figure*}

\section{The nebular properties of HaTr~5}
\label{sec:prop}

The VPHAS+ H$\alpha$+[N~{\sc ii}] image of HaTr\,5 in the top-right panel of Fig.~\ref{fig:img} \citep[see also Fig.~1 in][]
{Shara+2017} shows a mild elliptical $100\times92$ arcsec nebula with its major axis aligned along the Southeast-Northwest 
direction.  The nebula displays a limb-brightened morphology, enhanced along the 
Southeast direction, and prominent filaments across the nebula and blister-like "flocculent" structures.  
HaTr\,5 is surrounded by large-scale diffuse, patchy, H$\alpha$ emission. 
Extensive, patchy emission is also prominent in the mid-IR (bottom-right panel of Fig.~\ref{fig:img}).  
The nebula seems to be superimposed on a large-scale wide filament running from Northeast to Southwest.  
The optical and IR large scale emissions are apparently unrelated to HaTr\,5.

The one-dimensional SAAO and SOAR spectra of HaTr\,5 are shown in Fig.~\ref{fig:spec}.  
The two intermediate-dispersion SAAO spectra of HaTr\,5 display a limited number of emission lines typical of ionized nebulae, 
including H$\alpha$, [N~{\sc ii}] $\lambda\lambda$6548,6584, and the [S~{\sc ii}] $\lambda\lambda$6716,6731 doublet as listed in 
Table~\ref{tbl:spec_low}. 
As noted in Section 2.3, the [O~{\sc i}] $\lambda\lambda$6300,6363 emission lines seen in the 2024 spectra, but absent in the earlier 2014 SAAO spectra, are due to poor sky subtraction necessitated by using an offset sky exposure taken on a different night and are consequently deemed unreliable. 
No blue emission lines are detected in the 2014 spectrum, but the deeper 2024 spectrum detects weak [O~{\sc iii}] $\lambda$5007 and H$\beta$, and notably bright [O~{\sc ii}] $\lambda$3727, although the latter can be affected by uncertainties in the flux calibration at the blue end of the spectrum. 
The [O~{\sc iii}] line is weaker than H$\beta$ indicating a low ionization nebula. 
For comparison, Table~\ref{tbl:spec_low} also lists the emission lines and approximate line ratios detected 
in the deep SAAO SALT observations of the brightest Southeast nebular region reported by \citet{Shara+2017}. 
The exact values of the line ratios may differ notably among the different spectra.  
As for the differences observed for the low-excitation [N~{\sc ii}] $\lambda\lambda$6548,6584 and [S~{\sc ii}] 
$\lambda\lambda$6716,6731 emission lines, these can be attributed to the lower signal-to-noise ratio of the SAAO 2014 spectrum 
and to the differing nebular regions registered by the SAAO 2014 and 2024 spectra (East-West long-slit across the nebula center) and the SALT spectrum (long-slit along the bright Southeast nebular edge). 
The intensity of the [O~{\sc i}] $\lambda\lambda$6300,6363 doublet in the SALT spectrum confirms their excessive value in the SAAO 2024 spectrum.  

The observed H$\alpha$ to H$\beta$ ratio can be used to derive the logarithmic extinction coefficient $c\mathrm{(H}\beta\mathrm{)}$ using the expression 
\begin{equation}
    c(\mathrm{H}\beta) = \frac{1}{f(\mathrm{H}\alpha)} \log \frac{[F(\mathrm{H}\alpha)/F(\mathrm{H}\beta)]_\mathrm{observed}}{[F(\mathrm{H}\alpha)/F(\mathrm{H}\beta)]_\mathrm{theoretical}}, 
\end{equation}
where it will be adopted a theoretical H$\alpha$ to H$\beta$ flux ratio of 2.863 \citep[see][]{Luridiana2015} for recombination Case B, and a value of $f(\mathrm{H}\alpha)=-0.34$ according to the standard extinction law of \citet{Whitford1958}.  
The above expression then becomes 
\begin{equation}
    c(\mathrm{H}\beta) = \frac{1}{-0.34} \log \frac{[F(\mathrm{H}\alpha)/F(\mathrm{H}\beta)]_\mathrm{observed}}{2.863}, 
\end{equation}
which, for the observed H$\alpha$ to H$\beta$ flux ratios of 5.56 and 5.26 derived from the SpUpNIC and SALT spectra in Tab.~\ref{tbl:spec_low}, imply values of $c\mathrm{(H}\beta\mathrm{)}$ of 0.85 and 0.78, respectively, i.e., $A_\mathrm{V}$ of 1.9 and 1.7 mag for the canonical value of 3.1 for $R_V$.
An average value of $0.81\pm0.10$ will be hereafter adopted for $c\mathrm{(H}\beta\mathrm{)}$, where the uncertainty accounts both for the differences among these two spectra and for the H$\alpha$ to H$\beta$ line ratio uncertainties.  
The nebular electron density can be derived  from the line ratio of the [S~{\sc ii}] $\lambda\lambda$6716,6731 density-sensitive doublet line ratio \citep[see, for instance,][]{OF2006}. 
The [S~{\sc ii}] $\lambda$6716 to $\lambda$6731 line ratio derived from the SAAO 2004 SpUpNIC and SALT spectra, $\approx1.40$, is, however, below the linear [S~{\sc ii}] line ratio to electron density correlation range, in the so-called low-density limit.  
The electron density of the nebular shell can only be stated to be low, $\lesssim$100~cm$^{-3}$.

The high-dispersion SOAR GHTS 2-D background-subtracted spectra show emission in the H$\alpha$ and [N~{\sc ii}] $\lambda\lambda$6548,6584 lines, but the [O~{\sc i}] $\lambda\lambda$6300,6363 emission lines are not detected.
Since the SOAR spectra have much higher spectral resolution and dispersion than the SAAO SpUpNIC and SALT spectra, it demonstrates that the [O~{\sc i}] emission lines detected in these spectra may be affected by the poor removal at lower spectral resolution of the bright and highly variable [O~{\sc i}] night sky-lines.

\begin{figure}
\centering
\includegraphics[angle=0,width=0.975\linewidth]{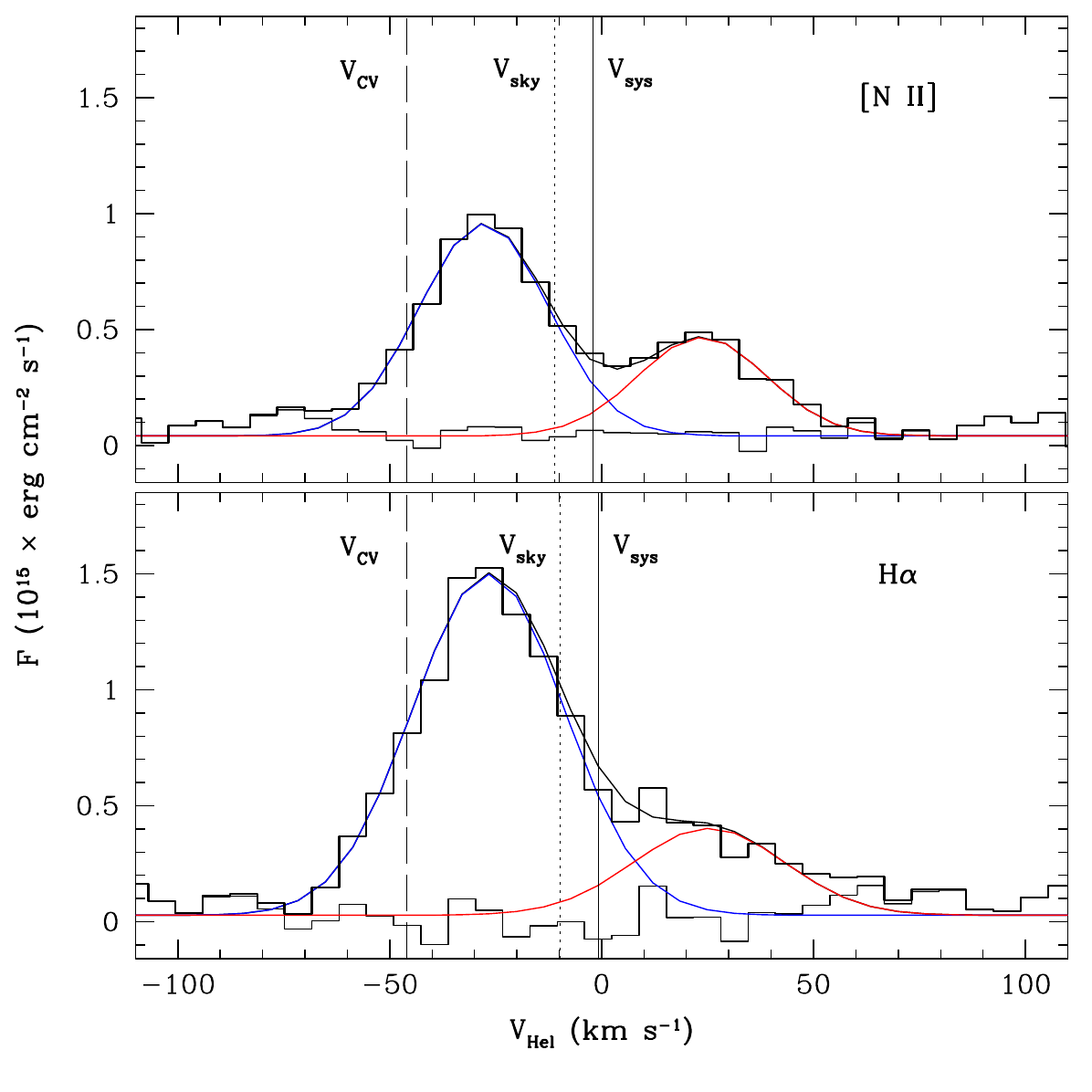}
\caption{\small 
SOAR GHTS one-dimensional spectrum (black thick histogram) with two-Gaussian fit of the [N~{\sc ii}] (top) and H$\alpha$ 
(bottom) emission lines (thin black line).  
Individual Gaussian curves are shown in blue and red and the residuals of the fit with a black thin histogram. 
The systemic velocity of HaTr\,5 and J17014-4306, and the velocity of the background nebular emission are marked by vertical solid ($V_\mathrm{sys}$), dashed ($V_\mathrm{CV}$), and dotted ($V_\mathrm{sky}$) lines, respectively. 
The spectral dispersion corresponds to $\simeq$6.4 km~s$^{-1}$ per pixel.
}
\label{fig:specfit}
\end{figure}

A closer inspection of the SOAR GHTS 2-D spectra reveals the typical lenticular shape of an expanding nebula both in the 
H$\alpha$ and [N~{\sc ii}] emission lines. The angular extent of the H$\alpha$ and [N~{\sc ii}]  emission undoubtedly associates 
it with HaTr\,5 (Fig.~\ref{fig:shape}), whereas the kinematically unresolved emission towards the Northeast can be most likely attributed to the large-scale diffuse emission. 
The double-component spectral profiles of the H$\alpha$ and [N~{\sc ii}] emission lines have been deblended using two-Gaussian fits (Fig.~\ref{fig:specfit}) to derive expansion and Heliocentric systemic velocities of $\simeq26$~km~s$^{-1}$ and $\simeq-1$~km~s$^{-1}$, respectively.  
The uncertainties of these velocities is estimated from the spectral profiles in Fig.~\ref{fig:specfit} to be smaller than half a pixel, i.e., $\pm$3~km~s$^{-1}$. 
For comparison, the Heliocentric velocity of the kinematically unresolved large-scale background emission around HaTr\,5 is 
$\simeq-10$~km~s$^{-1}$. The line widths ($\sigma$) of the H$\alpha$ and [N~{\sc ii}] $\lambda$6584 emission lines are 
18~km~s$^{-1}$ and 15~km~s$^{-1}$, respectively.

\begin{figure}
\centering
\includegraphics[angle=0,width=0.975\linewidth]{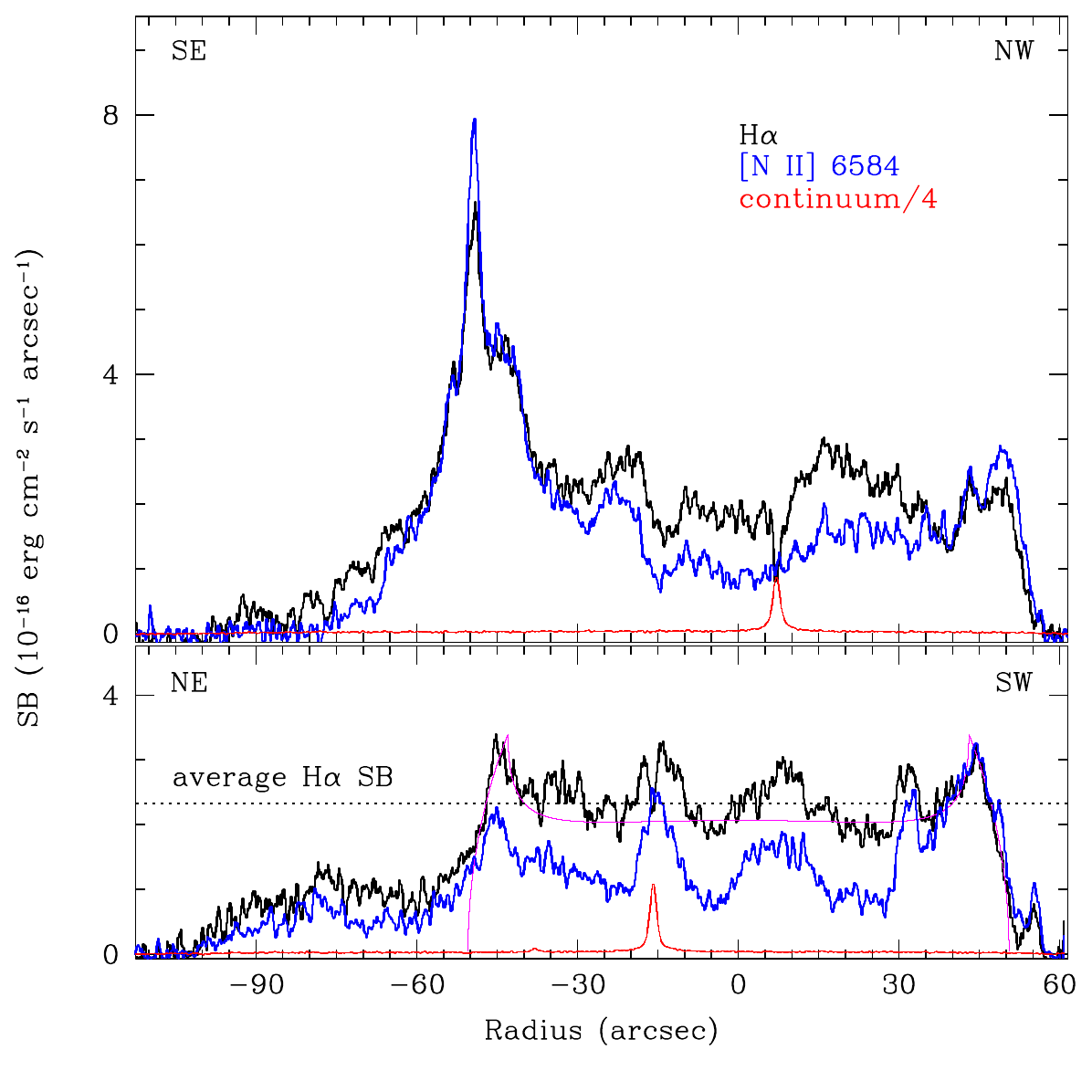}
\caption{\small 
Continuum-subtracted H$\alpha$ (black solid line) and [N~{\sc ii}] (blue solid line) surface brightness profiles of HaTr\,5 
along the SE-NW at PA=135$^\circ$ (top) and NE-SW PA=45$^\circ$ (bottom) directions extracted from the high-dispersion SOAR 
GOODMAN spectra.  The continuum emission scaled-down by a factor of four is shown in red, where the location of background stars 
registered by the slit provide accurate fiducial points. 
The horizontal thin-dotted line in the bottom panel marks the average value of the H$\alpha$ surface brightness. 
The magenta line in the bottom-panel is a synthetic H$\alpha$ emission line profile broadly describing the observed one (see text).
Note that the surface brightness of the [N~{\sc ii}] emission line is generally smaller than that of H$\alpha$, but at the Southeast and Southwest nebular edges. 
}
\label{fig:prof}
\end{figure}

The H$\alpha$ and [N~{\sc ii}] surface brightness (SB) radial profiles of HaTr\,5 presented in Fig.~\ref{fig:prof} are also 
consistent with a shell, with emission peaks along the Northeast-Southwest (minor axis) and Southeast-Northwest (major axis) 
directions consistent with the VPHAS+ H$\alpha$+[N~{\sc ii}] image of HaTr\,5.  
The H$\alpha$ SB is generally brighter than the [N~{\sc ii}] one, except for the edges of the shell major axis, very particularly 
the Southeast one, where the SB profiles have maximum values and [N~{\sc ii}] is brighter than H$\alpha$.
Otherwise the relatively high ``peak-to-center'' H$\alpha$ SB ratio in Fig.~\ref{fig:prof} is indicative of a thick shell.  
The surface brightness profile along the more representative Northeast-Southwest direction is roughly described by a thin outer shell with a thickness $\approx$15 per cent the nebular radius filled with material with  emissivity five times lower (magenta line in the bottom panel of Fig.~\ref{fig:prof}). 
The filling factor $\varepsilon$ of this multiple shell structure is $\simeq0.65$.

Assuming a constant velocity expansion, a kinematical age can be derived for HaTr\,5 from its angular radius, distance and expansion velocity.
An averaged angular radius of 50.5 arcsec is estimated from the size of the SB radial profiles at half-peak intensity. 
Thus a kinematical age of $(9200\pm1400) \times (D/\mathrm{1 kpc})$ yr is derived, where $D$ stands for the distance.  


Since kinematical information is available along two orthogonal long-slits covering the minor and major nebular axes (see Fig.~\ref{fig:img}), the 3-D physical structure of the expanding shell of HaTr\,5 can be mostly determined \citep[see, for instance,][]{Santamaria+2022} using the well-known spatio-kinematical modeling tool SHAPE \citep{Steffen2011}. 
This tool produces synthetic images and position-velocity (PV) diagrams for a given geometry (shape and orientation), expansion velocity law (a homologous expansion velocity is assumed here), and surface brightness distribution (including small-scale features such as filaments and knots, and asymmetric brightness distribution) of the shell.  
These are qualitatively compared to the observed ones, paying special attention to the nebula size, shape, and orientation in the image, to the expansion velocity and line tilt in the PN diagrams, and to small-scale features (clumps, filaments) in both the image and PV diagrams. 
The model is accordingly modified in an iterative process until a satisfactory match is reached such that any further parameter modification makes the synthetic image and PV diagrams depart from the observed ones. 
The best 3-D model for HaTr\,5 is an irregular oblate shell with a notorious dent, which was adjusted taking into account the line-tilt of the PV diagrams and their deviations with respect to a lenticular shape, and the filamentary appearance of the image and its deviations with respect to an elliptical morphology. 
This best model is shown as a mesh structure seen along different directions in Fig~\ref{fig:shape}-right.  
The SHAPE synthetic PV diagrams and image provide a reasonable representation of the observed PV diagrams and image in the left 
and middle panels of Fig~\ref{fig:shape}, respectively. The SHAPE model implies an expansion velocity of 27$\pm$3~km~s$^{-1}$, 
whereas the inclination angle with the line of sight of its major axis is estimated to be 18$^\circ\pm$2$^\circ$. 
The kinematical age of HaTr\,5 derived from this model is $(9300\pm1000) \times (D/\mathrm{1 kpc})$ yrs, consistent with that derived above.

The value of the H$\alpha$ flux of an ionized nebula is key to determine its ionized mass and root mean square (rms) density.  
These can be computed according to expressions 
\begin{equation}
M_\mathrm{i} = 0.035 \; \varepsilon^{1/2} \; \Theta^{3/2} \; D^{5/2} \; F_0(\mathrm{H}\alpha)^{1/2} \; M_\odot 
\label{eq:m}
\end{equation}
and 
\begin{equation}
n_\mathrm{e} = 1350 \; M_\mathrm{i} \; \left[ \frac{4 \pi}{3} \; (\Theta \; D)^3 \varepsilon \right]^{-1} \; \mathrm{cm}^{-3}
\label{eq:n}    
\end{equation}
where $F_0(\mathrm{H}\alpha)$ is the reddening corrected total nebular flux in units of $10^{-12}$ erg~cm$^{-2}$~s$^{-1}$ and 
$\Theta$ is the angular radius in arc\-min \citep{Pottasch1984}. 
The H$\alpha$ flux of HaTr\,5 can be computed from the SB radial profiles presented in Fig.~\ref{fig:prof}.  
These have been averaged radially to determine a SB($r$) function subsequently used to compute an average H$\alpha$ SB of $2.3\times10^{-16}$ 
erg~cm$^{-2}$~s$^{-1}$~arcsec$^{-2}$ and thus an observed H$\alpha$ 
flux\footnote{
For comparison the observed flux of the [N~{\sc ii}] $\lambda$6584 emission line is $2.0\times10^{-12}$ 
erg~cm$^{-2}$~s$^{-1}$, i.e., 0.85 times that of H$\alpha$. 
We note that \citet{Shara+2017}, by correcting the flux of their H$\alpha$+[N~{\sc ii}] image by the H$\alpha$/[N~{\sc ii}] ratio derived in the Southeast corner of HaTr\,5, where this ratio is minimal, inferred a lower H$\alpha$ flux, $\simeq1.0\times10^{-12}$ 
erg~cm$^{-2}$~s$^{-1}$.   } 
of $2.3\times10^{-12}$ erg~cm$^{-2}$~s$^{-1}$. 
The unreddened H$\alpha$ flux is then computed as 
\begin{equation}
    F_0(\mathrm{H}\alpha) = 10^{[(1-f(\lambda)) \, \times \, c(\mathrm{H}\beta)]} F(\mathrm{H}\alpha)
\end{equation}
that, for a $c(\mathrm{H}\beta)$ value of 0.81, implies an unreddened H$\alpha$ flux of $7.9\times10^{-12}$~erg~cm$^{-2}$~s$^{-1}$ (or $\log F_0(\mathrm{H}\alpha) = -11.10$).  
Therefore the ionized mass of HaTr\,5 is estimated to be $0.061 \times (D/\mathrm{1 kpc})^{5/2}$ M$_\odot$, with an rms density 
of $\simeq 50 \times (D/1~\mathrm{kpc})^{-1/2}$ cm$^{-3}$.

\section{Discussion}
\label{sec:disc}

The nebular properties of HaTr\,5 derived in the previous section allow assessing its true nature.  
The observed expansion of HaTr\,5 definitely discards it to be a Str\"omgren sphere around a hot WD, thus it remains to find out whether it is a nova remnant (Nova Sco\,1437) or a PN.

\subsection{HaTr\,5 as Nova Sco\,1437}

If HaTr\,5 were the nova remnant associated with Nova Sco\,1437, then its distance would be that of its putative progenitor J17014-4306.  
According to the latest Gaia data release DR3, the distance to J17014-4306 (Gaia source ID 5966150998003864960) is $988\pm37$ pc \citep[see, e.g.,][]{Bailer-Jones+2021}.  

\begin{figure*}
\centering
\includegraphics[bb=10 30 600 420,angle=0,width=0.98\linewidth]{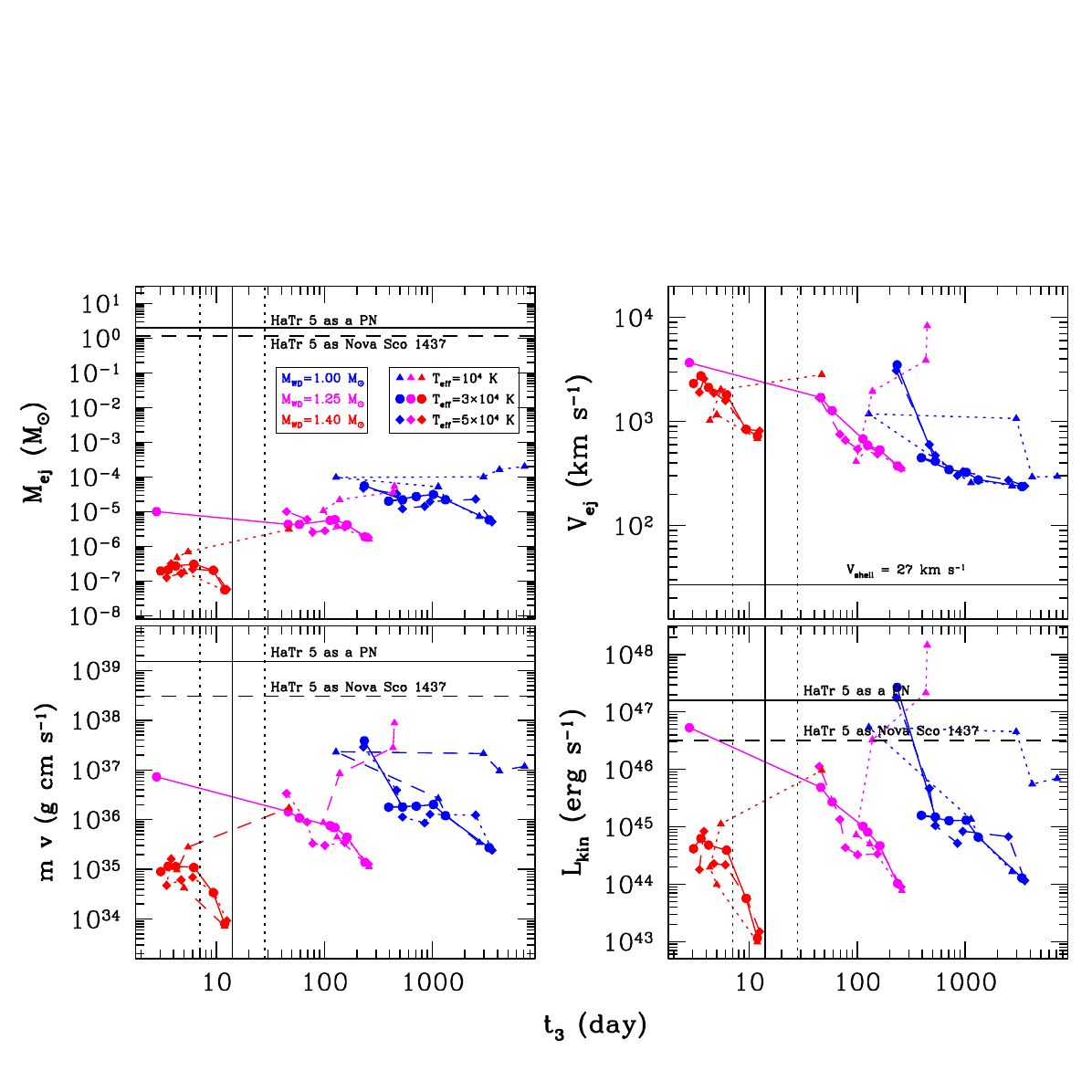}
\caption{\small 
Comparison of the mass (top-left), expansion velocity (top-right), linear momentum (bottom-left), and kinetic energy (bottom-right) of HaTr\,5 as a PN (horizontal solid line) and as Nova Sco\,1437 (horizontal dashed line) with theoretical expectations of CNe for 1.00 M$_\odot$, 1.25 M$_\odot$ (magenta) and 1.40 M$_\odot$ (red) WD of effective temperature 10,000, 30,000, and 50,000 K as labeled in the top-left panel \citep{Yaron+2005}.   
The vertical solid (dotted) line corresponds to the visibility time of 14 days (7 and 28 days) of Nova Sco\,1437 as registered in historical records. 
}
\label{fig:nova_model}
\end{figure*}

This distance implies a nebular ionized mass of $0.059\pm0.003$~M$_\odot$, a density $\simeq$50~cm$^{-3}$, and a physical radius of 0.24$\pm$0.01~pc. 
This mass is compared (top-left panel Fig.~\ref{fig:nova_model}) with theoretical predictions of mass ejected by CNe with WD mass of 1.00, 1.25, and 1.40~M$_\odot$ \citep{Yaron+2005}.  
Models with WD mass of 1.25 and 1.40~M$_\odot$ have the shortest values of $t_3$, which fit the short decline time of Nova Sco\,1437 of $\approx$14 days. 
This decline time, however, is questioned by the interpretation of historical Korean records \citep{Hoffmann2019}. 
Moreover the distance of 988 pc towards J17014-4306 implies a lower WD mass, $\simeq$1.0 \citep{Shara+2017}.   
Therefore models with WD mass of 1.00~M$_\odot$ have also been included in the comparison in the top-left panel Fig.~\ref{fig:nova_model}, whereas models with even lower WD mass, which have too large values of $t_3$, $\geq$438 days, will not be considered here.

This figure shows that the mass of HaTr\,5 at this distance is at least $10^4$ times larger than expected for a nova remnant.  
If it were assumed that the nova remnant has snow-plowed material from the ISM, the volume evacuated by HaTr\,5, corresponding to a mass $\approx 0.0029 \times n_{\rm ISM}$~M$_\odot$, would require a unusually high ISM density of $\approx$20~cm$^{-3}$.  
The expansion onto a high density, cold ISM can be expected to have swept up significant amounts of dust to make detectable its 
infrared emission, which is otherwise not detected in \emph{Spitzer} images.  
It must be noted, however, that shocks in nova ejecta may be very efficient in destroying dust grains \citep{Gehrz+1998}, but the present low expansion velocity of the shell questions this option. 

The expansion velocity of 27~km~s$^{-1}$ of HaTr\,5 also falls well below the expectation for a CN (top-right panel of Fig.~\ref{fig:nova_model}).  
\citet{Shara+2017} assumed its current expansion velocity to be 300~km~s$^{-1}$, which is much higher than the one in fact measured.  
They thus argued that the initial expansion velocity of the nova remnant had been slowed down by snow-plowing ISM material, but their adopted nova remnant deceleration rate, based on the conclusions of the analysis of the expansion of four nova remnants \citep{Duerbeck1987}, is questioned by the analysis of much higher quality data of three out of these four nova remnants, namely GK\,Per, V476\,Cyg, and DQ\,Her, actually supporting that they are in free expansion after having experienced negligible deceleration \citep{Liimets+2012,Santamaria+2020}.  
By assuming linear momentum conservation, the expansion velocity of a shell produced by the nova remnant depends only on the initial velocity of the ejecta and the ratio between the mass ejected and present mass.  
Thus the observed velocity and shell mass exceeds any possible combination of the ejecta mass and velocity for a CN (left-bottom panel of Fig.~\ref{fig:nova_model}).

The systemic velocity of HaTr\,5, $\simeq-1$~km~s$^{-1}$, also differs notably from that of J17014-4306, $-46$~km~s$^{-1}$ 
\citep{Shara+2017}.  On the assumption that HaTr\,5 consists mostly of ISM material swept-up by the ejecta of Nova Sco\,1437, 
its systemic velocity should match that of the ISM, but the  velocity of the nebular material along the line of sight, however, 
is also different, $\simeq-10$~km~s$^{-1}$.

Finally, the morphology of HaTr\,5, with a density enhancement towards the South-Southeast direction, is indicative of a 
possible interaction with the ISM of a nebula moving along this direction.  
The PA of this bow-shock-like feature, $\approx 125^\circ$, is mostly orthogonal to the proper motion of J17014-4306, which is 
aligned along PA $\simeq 203^\circ$.

\subsubsection{HaTr\,5 as Nova Sco\,1437 ... but closer}

The Gaia parallax (and thus the distance) to variable sources projected against diffuse nebular emission, as it is the case of CN progenitors, might be affected by systematic errors \citep[e.g.,][]{Schaefer+2018,Harvey+2020,Tappert+2020}.  
This does not seem to be the case for J17014-4306, whose Gaia DR3 parallax measurements are quite solid \citep{Gaia2023} and imply a distance consistent with that of $1030_{-30}^{+40}$ kpc derived by \citet{Schaefer2022}. 
Even then, let's consider a closer distance towards HaTr\,5, which would result in a smaller value of the ionized mass.

\begin{figure}
\centering
\includegraphics[bb=41 66 582 568,angle=0,width=0.975\linewidth]{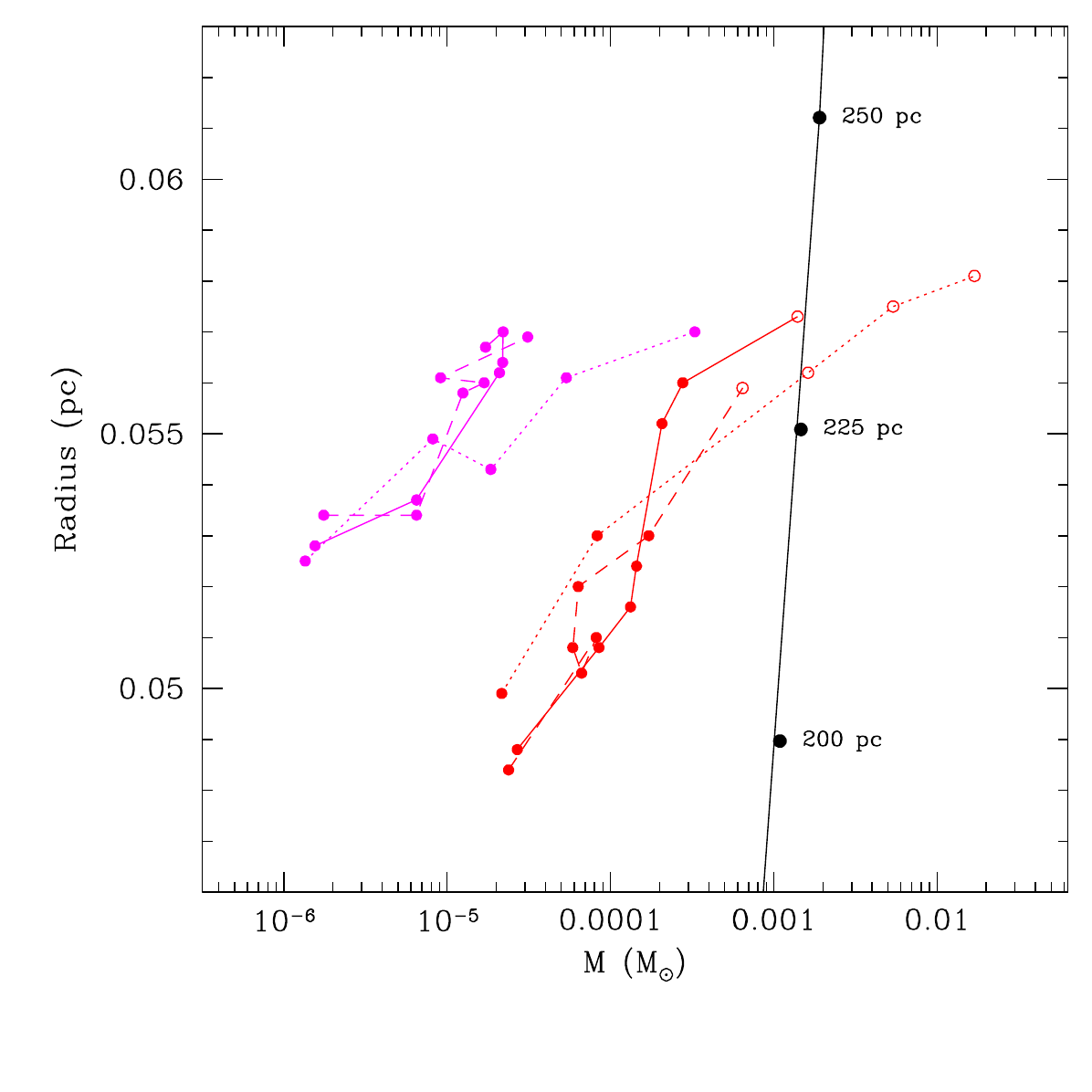}
\caption{\small 
Ionized mass and linear radius of nova remnants of theoretical CNe for 1.25 M$_\odot$ (magenta) and 1.40 M$_\odot$ (red) WDs of different temperatures \citep{Yaron+2005} as labeled in the top-left panel of Fig.~\ref{fig:nova_model}.  
The solid line corresponds to the ionized mass to distance relationship of $0.061 \times (D/\mathrm{1 kpc})^{5/2}$ M$_\odot$ derived for HaTr\,5 based on its intrinsic H$\alpha$ flux, angular diameter, and filling factor (Eq.~\ref{eq:m}). 
Models requiring ISM density above 20 cm$^{-3}$ are marked as open symbols. 
}
\label{fig:HaTr5_closer}
\end{figure}

Using the theoretical predictions on the initial ejecta mass and its average velocity \citep{Yaron+2005}, the current expansion velocity (27~km~s$^{-1}$) and age of the nova remnant ($\simeq$590 yrs) have been used to constrain the required ISM density to match these values.  
As a result, the nebular radius and mass of the nova remnant can be derived under the assumption of linear momentum conservation
\begin{equation}
    m_\mathrm{ej} \; v_\mathrm{ej} = \bigg( m_\mathrm{ej} + \frac{4}{3} \pi \rho_\mathrm{ISM} r^3 \bigg) \; v.
\end{equation}
These are compared with the nebular radius and ionized mass of HaTr\,5 at different distances in Fig.~\ref{fig:HaTr5_closer}, 
where models requiring unphysical ISM densities above 20 cm$^{-3}$ are marked as open symbols.

Depending on the input momentum according to the theoretical CN nova models of \citet{Yaron+2005}, the final nebular radius at the age of Nova Sco\,1437 is in a narrow range, 0.48--0.58 pc, but the swept-up ionized mass varies from $10^{-6}$ up to 0.02 M$_\odot$. 
There is no agreement between the predictions for CN remnants and the observed nebular radius and ionized mass unless the distance to the nova remnant is of the order of 200-250 pc, which is four to five times smaller than the Gaia distance to J17014-4306 and seems to be precluded by the small Gaia's distance error bar of 37 pc.  
Even then, the agreement is only reached at the expenses of an unphysically dense ISM.

\subsection{HaTr\,5 as a PN}

On the assumption that HaTr\,5 is a PN, its physical radius and subsequently its distance can be derived from the so-called 
reddening-corrected H$\alpha$ SB-radius relationship,
\begin{equation}
  \log \emph{SB}(\mathrm{H}\alpha) = -3.61(\pm0.11) \times \log R - 5.38(\pm0.09)
\label{eq:d}
\end{equation}
where $\emph{SB}(\mathrm{H}\alpha)$ is the reddening-corrected H$\alpha$ \emph{SB} in units of erg~cm$^{-2}$~s$^{-1}$~sr$^{-1}$ 
and $R$ is the nebular radius in pc \citep{Frew+2016}.  
Thus, given the reddening-corrected H$\alpha$ \emph{SB} of $3.5\times10^{-5}$ 
erg~cm$^{-2}$~s$^{-1}$~sr$^{-1}$, it results in a physical radius of $0.55\pm0.05$ pc  
according to Eq.~(\ref{eq:d}).  
The angular radius of 50.5 arcsec thus implies a distance of 2250$\pm$280 pc and, according to Eq.~(\ref{eq:m}), the mass of HaTr\,5 is 0.47~M$_\odot$, i.e., about 8 times larger than the mass at the distance of J17014-4306 and a very typical mass for a PN, while its rms density would be $\simeq$35~cm$^{-3}$.  
It must be noted that, although the volume occupied by HaTr\,5 at this distance is now significantly larger, the ISM mass snow-plowed, $\approx 0.02 \times n_{\rm ISM}$~M$_\odot$, would still be negligible for typical densities ($\approx$1 cm$^{-3}$) of the ISM.

These values compare relatively well with those presented by \citet{Frew+2016}, which derived a physical radius 25 per cent 
smaller (0.41 pc), a distance 35 per cent smaller (1470 pc), and an ionized mass 43 per cent smaller (0.27 $M_\odot$).  
The smaller physical radius, distance, and ionized mass are caused by their larger observed H$\alpha$ flux and adopted 
logarithmic extinction coefficient, which result in intrinsic H$\alpha$ flux and surface brightness about 3 times larger than 
those used here.

The larger extinction towards HaTr\,5 ($A_\mathrm{V}=1.7-1.9$ mag) than J17014-4306 \citep[$A_\mathrm{V}=1.1$ mag, as implied by the moderate reddening of its spectrum,][]{Shara+2017} also supports a larger distance for the nebula than for the CV.  
Indeed the reddening maps based on the inversion of color excess measurements provided by the STructuring by Inversion the Local Interstellar Medium \citep[Stilism,][]{Capitanio+2017} web site\footnote{\url{https://stilism.obspm.fr}} remarkably confirm the different distances proposed for the CV and the nebula:  
the extinction at the distance of J17014-4306 is $E(B-V)$ of $0.36\pm0.11$ mag (i.e., $A_V = 1.12\pm0.34$ mag), and it increases to $0.59\pm0.10$ mag (i.e., $A_V = 1.83\pm0.31$ mag) at 2085 pc, which is the maximum distance probed by Stilism along this direction, but within the uncertainty range of the PN distance of $2250\pm280$ pc.  

\subsubsection{On the nebular spectrum of HaTr\,5}

The previous classification of HaTr\,5 as a ``possible PN'' by \citet{Parker+2016} was motivated by the large [S~{\sc ii}] to 
H$\alpha$ line ratio of its nebular spectrum and by the absence of a central star.  
Note that the spectroscopic investigation of large, low surface brightness nebulae is problematic, further hampered in the case 
of HaTr\,5 by the difficult background subtraction caused by the diffuse emission along the line of sight.  
Indeed \citet{Parker+2016} claimed a [S~{\sc ii}] 6717+6731/H$\alpha$ line ratio of 1.04 based on the 2014 SAAO spectrum, but the more recent 2024 SAAO spectrum reduces this ratio to 0.63, which is further brought down to 0.52 by 
\citet{Shara+2017}.  
The line ratios of HaTr\,5 seem typical of supernova remnants, but the comprehensive set of photoionization simulations provided by the Mexican Million Models\footnote{https://sites.google.com/site/mexicanmillionmodels} database \citep{Morisset+2015} shows it is also feasible for a PN to occupy this locus on diagnostic diagrams \citep[see, for instance,][]{Belfiore+2024}, 
Otherwise the central star of an old PN can be expected to have a low luminosity \citep{MB2016}, making it undetectable.

To model the nebular spectral properties of HaTr\,5 and to derive key parameters characterizing the properties of its central 
star, we have used the photoionization code Cloudy \citep[version 23.0;][]{Chatzikos+2023}.  
The model adopts a spherical shell geometry with two concentric shells, with an inner empty cavity. The shells are defined by 
three characteristic radii: an inner radius of $r_\mathrm{in} = 0.12$\ pc, a middle radius of $r_\mathrm{mid} = 0.35$ pc, and 
an outer radius of $r_\mathrm{out} = 0.55$ pc. 
The inner and outer shells have number densities $n_\mathrm{1}=15$ cm$^{-3}$ and $n_\mathrm{2}=60$ cm$^{-3}$, respectively. 
The chemical abundances of the model were set to those reported by \citet{Aller1983} for PNe, with increased S and N abundances 
by 10 and 13 per cent, respectively, and O abundances decreased by 15 percent. 
We used a blackbody spectrum for the incident radiation, varying the effective temperature ($T_\mathrm{eff}$) and stellar luminosity ($L_\star$) according to the stellar evolution models by \citet{MB2016}. 
The model emission line ratios were then used to constrain the model parameters, iteratively adjusting them to minimize the 
differences between observed and synthetic spectra.

The expected line ratios of the best-fitting model produce a noticeable agreement with the observed spectra (see Tab.~\ref{tbl:spec_low}).  
The parameters of this best-fitting model correspond to a blackbody spectrum with effective temperature of 93,000~K ($\log T_\mathrm{eff} = 4.97$) and luminosity of 46 $L_\odot$ ($\log (L_\star/\mathrm{L}_\odot) = 1.66$). 
According to the evolutionary tracks of \citet{MB2016}, the parameters of such star are consistent with a PN central star having a remnant mass of approximately 0.58 M$_\odot$. 
The corresponding age of the star, inferred from these tracks, is greater than 10,000 years. 
The atypically low [O~{\sc iii}] $\lambda$5007 to H$\alpha$ line ratio, but high [O~{\sc ii}] $\lambda$3727 to [O~{\sc iii}] $\lambda$5007 and [N~{\sc ii}] and [S~{\sc ii}] to H$\alpha$ line ratios are most likely indicative of recombination processes caused by the decline of the ionizing flux, as seen in recombining haloes of PNe \citep{Corradi+2000}.

\subsubsection{On the absence of a central star for HaTr\,5}

The spectral fitting of HaTr\,5 implies an evolved central star.  
A late PN central star is also supported by the "relaxed'' shell structure implied by its considerable thickness \citep{JSS2013}, which reveals a low internal thermal pressure as expected for an evolved central star whose stellar wind mass-loss rate is diminished.  
At the distance of 2250 pc, the kinematic age of HaTr\,5 is 20,000$\pm$2,300 yr.  
This kinematic age can only be considered indicative of an evolved PN, as the kinematic age of a PN can be very different from its ``real'' age (the time since its progenitor entered the post-Asymptotic Giant Branch phase).  
Actually, for an evolved PN with a central star with $T_\mathrm{eff}$ of 93,000 K, hydrodynamical simulations show that the real age is in the range 0.5--1.0 times the kinematic age, depending on the initial mass of the progenitor and envelope structure \citep[Fig.~4 in][]{Schonberner+2014}. 
Thus the nebular kinematic age implies a true age in the range 10,000 to 20,000 yr, consistent with the age of the progenitor star.

\begin{table}
\caption{Stars projected onto HaTr\,5 detected in the VPHAS+ $u$ image.}
\label{tbl:vphas_phot}
\centering
\begin{tabular}{lccccc}
\hline\hline
\multicolumn{1}{c}{Star$^a$} &  
\multicolumn{1}{c}{$\alpha$} &
\multicolumn{1}{c}{$\delta$} &
\multicolumn{1}{c}{$u$} &
\multicolumn{1}{c}{$g$} &  
\multicolumn{1}{c}{($u-g$)} \\ 
\multicolumn{1}{l}{} & 
\multicolumn{2}{c}{(J2000)} & 
\multicolumn{1}{c}{(mag)} & 
\multicolumn{1}{c}{(mag)} & 
\multicolumn{1}{c}{(mag)} \\
\hline
\noindent
\#1     & 17:01:28.07 & $-43$:05:55.3 & $\dots$ & 19.18 & $\dots$ \\
\#2     & 17:01:29.06 & $-43$:05:31.5 & 19.02  & 17.25 & $+1.77$ \\
\#3     & 17:01:28.29 & $-43$:05:40.4 & 18.80  & 17.77 & $+1.03$ \\
\#4$^b$ & 17:01:28.15 & $-43$:06:12.4 & 16.24  & 16.41 & $-0.17$ \\
\hline
\end{tabular}
\tablefoot{$^a$ Stars are labeled on the $u$ band VPHAS+ image in the top-left panel of Fig.~\ref{fig:img}. $^b$ J17014-4306.}
\end{table}

Using the parameters derived from the Cloudy model, namely $\log(T_\mathrm{eff}) = 4.97$ and $\log(L/\mathrm{L}_\odot) = 
1.66$, and its distance, the expected $u$ and $g$ magnitudes of the central star ionizing HaTr\,5 can be computed. 
For this, the blackbody model adopted for the star was integrated over the respective OmegaCAM filter transmission curves as 
obtained from the publicly accessible page SVO Filter Profile Service "Carlos Rodrigo"{\footnote{\url{http://svo2.cab.inta-
csic.es/svo/theory/fps3/index.php?mode=browse}}}.  
Thus, at the distance towards HaTr\,5 its central star would have $u$ and $g$ magnitudes of 30.2 and 31.4 mag, respectively. 
These values are much larger than the limiting magnitudes of VPHAS+ in the $u$ and $g$ bands of 21.8 and 22.6 mag, respectively.  

Consequently, our photoionization Cloudy model predicts that the central star of HaTr\,5 is way too 
faint to be detected in the VPAHS+ $u$ and $g$ images.  
Indeed the four stars projected onto HaTr\,5 detected in the VPHAS+ $u$-SDSS image, as marked in the top-left panel of 
Fig.~\ref{fig:img}, have $u$-SDSS and $g$-SDSS magnitudes (Tab.~\ref{tbl:vphas_phot}) much smaller than those expected from the central star of HaTr\,5.  
The brightest and bluest star, marked \#4, is J17014-4306, which has an intrinsic $u-g$ color of $-0.70$ mag, after an extinction correction $A_\mathrm{V} = 1.8$ mag is applied. 
The closest star to the nebula center\footnote{\citet{Shara+2017} determined the nebula center to be at $\alpha =$ 17:01:28.55 
and $\delta =$ --43:05:59.3 (J2000). }, about 6.3 arcsec Northwest, marked \#3, is the Gaia source ID~5966150998003866368. 
This was selected as the very likely central star of HaTr\,5 by \citet{Gonzalez+2021}, with an intrinsic $(G_\mathrm{BP}-
G_\mathrm{RP})_\circ$ of --2.16 mag below their threshold of --0.2 mag for an ionizing source.  
It must be noted, however, that the value of the extinction correction used in that work ($A_\mathrm{V} = 27.9$ mag), computed from Gaia's photometric measurements, differs notably from the value derived here for the nebula. 
After the extinction correction $A_\mathrm{V} = 1.8$ mag of the nebula is applied, the intrinsic $(u-g)_\circ$ and $(G_\mathrm{BP}-G_\mathrm{RP})_\circ$ colors would be +0.50 mag and +1.0 mag, respectively, which are not consistent with the central star of a PN \citep[unless the presence of a companion star were to be claimed,][]{Augusteijn+2008}, but with a star with the effective temperature $\simeq$6630~K provided by the Gaia database.   
In summary, none of the stars detected in the present VPHAS+ $u$-SDSS band image can be associated with HaTr\,5, which is 
consistent with a low-luminosity PN central star.


\section{Final remarks}
\label{sec:remarks}

\citet{Shara+2017} had argued that the CV J17014-4306 is the historical Nova Sco\,1437 and that the nebula HaTr\,5 is the 
associated nova remnant based on their proper-motion measurements of J17014-4306 ($\mu_\alpha = -12.7\pm1.8$ mas~yr$^{-1}$, 
$\mu_\delta = -27.7\pm1.2$ mas~yr$^{-1}$), which placed it close to the center of HaTr\,5 by the time of the nova event.  
Recent more accurate Gaia DR3 proper-motion measurements of this star (source ID 5966150998003864960), however, reduced them by 
a factor of 3 ($\mu_\alpha = -4.14\pm0.05$ mas~yr$^{-1}$, $\mu_\delta = -9.701\pm0.036$ mas~yr$^{-1}$), placing J17014-4306 
offset from the center of HaTr\,5 by the time of Nova Sco\,1437 eruption.  
Although this correction does not really seem relevant \citep[see, for instance,][]{Hoffmann2019}, it adds to the different 
radial systemic velocities of HaTr\,5 ($-1$ km~s$^{-1}$) and J17014-4306 ($-46$ km~s$^{-1}$), and to the almost orthogonal PAs 
of the proper-motion of J17014-4306 (PA $\simeq 203^\circ$) and orientation of the intensity enhancement of the nebular 
emission of HaTr\,5 (PA $\approx 125^\circ$).

Most of the nebular properties of HaTr\,5 presented here support it to be a textbook evolved PN.  
Particularly its ionized mass is much larger than expected for a nova remnant, while its expansion velocity ($\lesssim 30$ 
km~s$^{-1}$) is much smaller.  Models of CNe appropriate for the speed class of Nova Sco\,1437 fail to provide the linear 
momentum of the expanding nebula.  Incidentally, it must be noted that J17014-4306 would have been $\approx3$ arcmin away 
from the nebula center when the PN formed, making it very unlikely its WD to have been the progenitor of this PN.

HaTr\,5 had been in the past conservatively classified as a possible PN \citep{Parker+2016} given that its central star is not 
detected in available photometric surveys and that its optical spectrum displays a notably large value of the [S~{\sc ii}] to 
H$\alpha$ line ratio (in the range $\approx1.0$ to $\approx0.5$), which is suggestive of shocks as found in supernova or nova 
remnants.  The analysis of its spectrum using a Cloudy model, however, supports that the observed line ratio is consistent with 
an old evolved PN whose central star is faint and hot.  
Such star would be undetectable at the distance towards HaTr\,5 of 2.25 kpc under the assumption it is a PN.

It is concluded that the nebular properties of HaTr\,5 are consistent with those of an old, evolved PN rather than a nova 
remnant. 
Its association with the CV J17014-4306 is spurious and thus HaTr\,5 cannot be identified with the remnant of Nova Sco\,1437, neither it provides any constraint to the transition time from nova-like binaries to dwarf novae proposed by \citet{Shara+2017}.


\begin{acknowledgements}

The authors acknowledge the helpful comments and suggestion of an anonymous referee that helped greatly to improve the scientific content of this manuscript. 
M.A.G.\ acknowledges financial support from grants CEX2021-001131-S funded by MCIN/AEI/10.13039/501100011033 and PID2022-142925NB-I00 from the Spanish Ministerio de Ciencia, Innovaci\'on y Universidades (MCIU) cofunded with FEDER funds. 
E.S.\ is funded by UNAM DGAPA postdoctoral fellowship. 
G.L.\ acknowledges CAPES (proc.\ 88887.948597/2024-00).
E.S., J.B.R.G., and J.A.T.\ thank support by UNAM PAPIIT project IN102324. 
Q.A.P.\ thanks the Hong Kong Research Grants Council for GRF research support under grants 17326116 and 17300417. 
D.R.G.\ acknowledges FAPERJ (E-26/200.527/2023) and CNPq (315307/2023-4) grants. 
A.R.\ and H.Y.\ thanks HKU and QAP for provision of postdoctoral and PhD scholarship funds from RMGS grant awarded to the LSR. \\
This research has made use of the software ASTROPY \citep{Astropy+2013,Astropy+2018}, Cloudy \citep{Ferland+2013}, and Source Extractor \citep{BA1996}, and an extensive use of NASA's Astrophysics Data System (ADS).
This research has made use of the NASA/IPAC Infrared Science Archive, which is funded by the National Aeronautics and Space Administration and operated by the California Institute of Technology.
This work is based 
on archival data obtained with the Spitzer Space Telescope, which was operated by the Jet Propulsion Laboratory, California Institute of Technology under a contract with NASA. Support for this work was provided by NASA, 
on observations obtained at the Southern Astrophysical Research (SOAR) telescope, which is a joint project of the Minist\'erio da Ciencia, Tecnologia, e Inova\c c\~ao (MCTI) da Rep\'ublica Federativa do Brasil, the U.S.\ National Optical Astronomy Observatory (NOAO), the University of North Carolina at Chapel Hill (UNC), and Michigan State University (MSU), 
on data products from observations made with ESO Telescopes at the La Silla Paranal Observatory under programme ID 177.D-3023, as part of the VST Photometric H$\alpha$ Survey of the Southern Galactic Plane and Bulge (VPHAS+), 
on data products of the SuperCOSMOS H-alpha Survey (SHS), 
and 
on observations made with the CCDSPEC, SpUpNIC, and SALT instruments at the South African Astronomical Observatory (SAAO). 

\end{acknowledgements}



\begin{thebibliography}{}

\bibitem[Aller \& Czyzak(1983)]{Aller1983} Aller, L.~H. \& Czyzak, S.~J.\ 1983, \apjs, 51, 211. 


\bibitem[Astropy Collaboration et al.(2013)]{Astropy+2013}
Astropy Collaboration, Robitaille, T. P., Tollerud, E. J., et al.\ 2013, \aap, 558, A33.

\bibitem[Astropy Collaboration et al.(2018)]{Astropy+2018}
Astropy Collaboration, Price-Whelan, A. M., Sipocz, B. M., et al. 2018, \aj, 156, 123..

\bibitem[Augusteijn et al.(2008)]{Augusteijn+2008} 
Augusteijn, T., Greimel, R., van den Besselaar, E.~J.~M., et al.\ 2008, \aap, 486, 843. 

\bibitem[Bailer-Jones et al.(2021)]{Bailer-Jones+2021} 
Bailer-Jones, C.~A.~L., Rybizki, J., Fouesneau, M., et al.\ 2021, \aj, 161, 147 



\bibitem[Belfiore et al.(2024)]{Belfiore+2024} 
Belfiore, F., Ginolfi, M., Guillermo Blanc, G.\ et al.\ 2024, \aaps, in press, arXiv:2410.16370
 
\bibitem[Bertin \& Arnouts(1996)]{BA1996} 
Bertin, E. \& Arnouts, S.\ 1996, \aaps, 117, 393. 

\bibitem[Bond \& Miszalski(2018)]{BM2018} 
Bond, H.~E. \& Miszalski, B.\ 2018, \pasp, 130, 094201. 

\bibitem[Capitanio et al.(2017)]{Capitanio+2017} 
Capitanio, L., Lallement, R., Vergely, J.~L., et al.\ 2017, \aap, 606, A65. 

\bibitem[Chatzikos et al.(2023)]{Chatzikos+2023} 
Chatzikos, M., Bianchi, S., Camilloni, F., et al.\ 2023, \rmxaa, 59, 327. 

\bibitem[Corradi et al.(2000)]{Corradi+2000} 
Corradi, R.~L.~M., Sch{\"o}nberner, D., Steffen, M., et al.\ 2000, \aap, 354, 1071

\bibitem[Crause et al.(2019)]{Crause+2019} 
Crause, L.~A., Gilbank, D., Gend, C. van ., et al.\ 2019, Journal of Astronomical Telescopes, Instruments, and Systems, 5, 024007. 

\bibitem[Dilday et al.(2012)]{Dilday+2012} 
Dilday, B., Howell, D.~A., Cenko, S.~B., et al.\ 2012, Science, 337, 942. 

\bibitem[Drew et al.(2014)]{Drew+2014} 
Drew, J.~E., Gonzalez-Solares, E., Greimel, R., et al.\ 2014, \mnras, 440, 2036. 

\bibitem[Duerbeck(1987)]{Duerbeck1987} 
Duerbeck, H.~W.\ 1987, \apss, 131, 461. 

\bibitem[Ferland et al.(2013)]{Ferland+2013} 
Ferland, G.~J., Porter, R.~L., van Hoof, P.~A.~M., et al.\ 2013, \rmxaa, 49, 137. 

\bibitem[Frew \& Parker(2010)]{FP2010} 
Frew, D.~J. \& Parker, Q.~A.\ 2010, \pasa, 27, 129. 

\bibitem[Frew et al.(2016)]{Frew+2016} 
Frew, D.~J., Parker, Q.~A., \& Boji{\v{c}}i{\'c}, I.~S.\ 2016, \mnras, 455, 1459. 

\bibitem[Gaia Collaboration et al.(2023)]{Gaia2023} 
Gaia Collaboration, Vallenari, A., Brown, A.~G.~A., et al.\ 2023, \aap, 674, A1. 

\bibitem[Gehrz et al.(1998)]{Gehrz+1998} 
Gehrz, R.~D., Truran, J.~W., Williams, R.~E., et al.\ 1998, \pasp, 110, 3. 

\bibitem[Gonz{\'a}lez-Santamar{\'\i}a et al.(2021)]{Gonzalez+2021} 
Gonz{\'a}lez-Santamar{\'\i}a, I., Manteiga, M., Manchado, A., et al.\ 2021, \aap, 656, A51. 

\bibitem[Greenstein(1984)]{Greenstein1984} 
Greenstein, J.~L.\ 1984, \apj, 276, 602. 

\bibitem[Hartl \& Tritton(1985)]{HaTr1985} 
Hartl, H. \& Tritton, S.~B.\ 1985, \aap, 145, 41

\bibitem[Harvey et al.(2020)]{Harvey+2020} 
Harvey, E.~J., Redman, M.~P., Boumis, P., et al.\ 2020, \mnras, 499, 2959. 

\bibitem[Hoffmann(2019)]{Hoffmann2019} 
Hoffmann, S.~M.\ 2019, \mnras, 490, 4194. 

\bibitem[Honeycutt et al.(2011)]{HRK2011} 
Honeycutt, R.~K., Robertson, J.~W., \& Kafka, S.\ 2011, \aj, 141, 121. 

\bibitem[Jacob et al.(2013)]{JSS2013} 
Jacob, R., Sch{\"o}nberner, D., \& Steffen, M.\ 2013, \aap, 558, A78. 

\bibitem[Liimets et al.(2012)]{Liimets+2012} 
Liimets, T., Corradi, R.~L.~M., Santander-Garc{\'\i}a, M., et al.\ 2012, \apj, 761, 34. 

\bibitem[Luridiana et al.(2015)]{Luridiana2015} Luridiana, V., Morisset, C., \& Shaw, R.~A.\ 2015, \aap, 573, A42

\bibitem[Miller Bertolami(2016)]{MB2016} 
Miller Bertolami, M.~M.\ 2016, \aap, 588, A25. 

\bibitem[Miszalski et al.(2016)]{Miszalski+2016} 
Miszalski, B., Woudt, P.~A., Littlefair, S.~P., et al.\ 2016, \mnras, 456, 633. 

\bibitem[Morisset et al.(2015)]{Morisset+2015} 
Morisset, C., Delgado-Inglada, G., \& Flores-Fajardo, N.\ 2015, \rmxaa, 51, 103. 

\bibitem[Mr{\'o}z et al.(2016)]{Mroz+2016} 
Mr{\'o}z, P., Udalski, A., Pietrukowicz, P., et al.\ 2016, \nat, 537, 649. 

\bibitem[Osterbrock \& Ferland(2006)]{OF2006} 
Osterbrock, D.~E. \& Ferland, G.~J.\ 2006, Astrophysics of gaseous nebulae and active galactic nuclei, 2nd. ed. by D.E. Osterbrock and G.J. Ferland. Sausalito, CA: University Science Books, 2006

\bibitem[Parker(2022)]{Parker2022} 
Parker, Q.~A.\ 2022, Frontiers in Astronomy and Space Sciences, 9, 895287. 

\bibitem[Parker et al.(2016)]{Parker+2016} 
Parker, Q.~A., Boji{\v{c}}i{\'c}, I.~S., \& Frew, D.~J.\ 2016, Journal of Physics Conference Series, 728, 032008. 

\bibitem[Parker et al.(2005)]{Parker+2005} 
Parker, Q.~A., Phillipps, S., Pierce, M.~J., et al.\ 2005, \mnras, 362, 689. 

\bibitem[Pottasch(1984)]{Pottasch1984} 
Pottasch, S.~R.\ 1984, Astrophysics and Space Science Library, Dordrecht: Reidel, 1984. 

\bibitem[Sahman et al.(2018)]{Sahman+2018} 
Sahman, D.~I., Dhillon, V.~S., Littlefair, S.~P., et al.\ 2018, \mnras, 477, 4483. 

\bibitem[Santamar{\'\i}a et al.(2020)]{Santamaria+2020} 
Santamar{\'\i}a, E., Guerrero, M.~A., Ramos-Larios, G., et al.\ 2020, \apj, 892, 60. 

\bibitem[Santamar{\'\i}a et al.(2022)]{Santamaria+2022} 
Santamar{\'\i}a, E., Guerrero, M.~A., Zavala, S., et al.\ 2022, \mnras, 512, 2003. 

\bibitem[Schaefer(2018)]{Schaefer+2018} 
Schaefer, B.~E.\ 2018, \mnras, 481, 3033. 

\bibitem[Schaefer(2022)]{Schaefer2022} 
Schaefer, B.~E.\ 2022, \mnras, 517, 6150. 

\bibitem[Sch{\"o}nberner et al.(2014)]{Schonberner+2014} 
Sch{\"o}nberner, D., Jacob, R., Lehmann, H., et al.\ 2014, Astronomische Nachrichten, 335, 378. 

\bibitem[Shara(1984)]{Shara1984} 
Shara, M.\ 1984, \jrasc, 78, 166

\bibitem[Shara et al.(2017)]{Shara+2017} 
Shara, M.~M., I{\l}kiewicz, K., Miko{\l}ajewska, J., et al.\ 2017, \nat, 548, 558. 

\bibitem[Shara et al.(1986)]{Shara+1986} 
Shara, M.~M., Livio, M., Moffat, A.~F.~J., et al.\ 1986, \apj, 311, 163. 

\bibitem[Shara et al.(2007)]{Shara+2007} 
Shara, M.~M., Martin, C.~D., Seibert, M., et al.\ 2007, \nat, 446, 159. 

\bibitem[{\v{S}}imon(2023)]{Simon2023} 
{\v{S}}imon, V.\ 2023, \aj, 165, 102. 

\bibitem[Somers et al.(1996)]{Somers+1996} 
Somers, M.~W., Mukai, K., \& Naylor, T.\ 1996, \mnras, 278, 845. 

\bibitem[Steffen et al.(2011)]{Steffen2011} Steffen, W., Koning, N., Wenger, S., et al.\ 2011, IEEE Transactions on Visualization and Computer Graphics, 17, 454

\bibitem[Tappert et al.(2012)]{Tappert+2012} 
Tappert, C., Ederoclite, A., Mennickent, R.~E., et al.\ 2012, \mnras, 423, 2476. 

\bibitem[Tappert et al.(2020)]{Tappert+2020} 
Tappert, C., Vogt, N., Ederoclite, A., et al.\ 2020, \aap, 641, A122. 

\bibitem[Tody(1993)]{Tody1993} 
Tody, D.\ 1993, Astronomical Data Analysis Software and Systems II, 52, 173

\bibitem[van Dokkum(2001)]{vanDokkum2001} 
van Dokkum, P.~G.\ 2001, \pasp, 113, 1420. 

\bibitem[Vogt et al.(2018)]{Vogt+2018} 
Vogt, N., Tappert, C., Puebla, E.~C., et al.\ 2018, \mnras, 478, 5427. 

\bibitem[Whitford(1958)]{Whitford1958} Whitford, A.~E.\ 1958, \aj, 63, 201

\bibitem[Yaron et al.(2005)]{Yaron+2005} 
Yaron, O., Prialnik, D., Shara, M.~M., et al.\ 2005, \apj, 623, 398. 


\end{thebibliography}
\end{document}